\documentclass[aps,prc,twocolumn,floatfix,superscriptaddress,nofootinbib]{revtex4-1}
\usepackage{graphicx}
\usepackage[pdftex,colorlinks,citecolor=blue,bookmarks]{hyperref}
\usepackage{bm}
\usepackage{amsmath}
\usepackage{longtable}

\begin{document}
\title{Variational approximations to exact solutions in shell-model valence spaces: \\ systematic calculations in the $sd$-shell}
\author{Adri\'an S\'anchez-Fern\'andez} 
\address{Departamento de F\'isica Te\'orica, Universidad Aut\'onoma de Madrid, E-28049 Madrid, Spain}
\author{Benjamin Bally} 
\address{Departamento de F\'isica Te\'orica, Universidad Aut\'onoma de Madrid, E-28049 Madrid, Spain}
\author{Tom\'as R. Rodr\'iguez} 
\address{Departamento de F\'isica Te\'orica, Universidad Aut\'onoma de Madrid, E-28049 Madrid, Spain}
\address{Centro de Investigaci\'on Avanzada en F\'isica Fundamental-CIAFF-UAM, E-28049 Madrid, Spain}
\begin{abstract}

We study the ability of variational approaches based on self-consistent mean-field and beyond-mean-field methods to reproduce exact energies and electromagnetic properties of the nuclei defined within the $sd$-shell valence space using the non-trivial USD Hamiltonian. In particular, Hartree-Fock-Bogoliubov (HFB), variation after particle-number projection (VAPNP) and projected generator coordinate methods (PGCM) are compared to exact solutions {obtained by} the full diagonalization of the Hamiltonian. We analyze the role played by the proton-neutron ($pn$) mixing as well as the quadrupole and pairing degrees of freedom (including both isoscalar and isovector channels) in the description of the spectra of even-even, even-odd and odd-odd nuclei in the whole $sd$-shell. 

\end{abstract}
\maketitle
\section{Introduction}
Self-consistent mean-field methods and their beyond-mean-field extensions are among the most successful techniques to study the structure of the atomic nucleus from a microscopic point of view. They rely on the Ritz variational principle to approach the exact solution of the nuclear quantum many-body problem. Traditionally, these methods have been mostly used in the context of  phenomenological energy density functionals and were applied to the study of various nuclear phenomena~\cite{Bender03a,NIKSIC2011519,Egido_2016,Robledo_2018}. More recently, these methods 
have been considered as numerically convenient and physically sound approximations to the exact diagonalization in valence-space calculations \cite{Gao15a,Jiao17a,Bally19a,Shimizu21a} as well as a powerful means to include important static correlations in the reference states of nuclear \emph{ab initio} calculations \cite{Yao18a,Yao20a,Frosini21a}.

The usual starting point of these methods is the definition of a variational space made of product-like many-body wave functions such as the Slater determinants, or better, the Bogoliubov quasiparticle vacua \cite{RS80a}. 
When using the latter, the matrix elements of the underlying Bogoliubov transformations are used as variational parameters
that are determined by solving the Hartree-Fock-Bogoliubov (HFB) equations, possibly with constraints. In general, the lowest variational energy corresponds to a deformed product state, i.e.\ a product state that breaks one or several symmetries of the nuclear Hamiltonian. While not physical in a finite system such as the nucleus, this spontaneous symmetry breaking phenomenon is of great advantage to take into account important static correlations (e.g.\ pairing or multipole deformations) while conserving the simplicity of product states as working many-body wave functions.

Next steps are subsequently taken to improve the wave functions of the system such as symmetry restorations and configuration mixing. One of the most sophisticated techniques of this kind is the so-called projected generator coordinate method (PGCM) \cite{RS80a,Hill53a,Griffin57a} where the trial wave functions are defined as linear combinations of different symmetry-projected quasiparticle vacua. These vacua are usually generated by solving multiple times the constrained HFB equations and using the values of the constraints as generator coordinates. The quality of the PGCM method with respect to the exact solution of the nuclear Hamiltonian depends 
on: i) the generality of the Bogoliubov transformations defining the quasiparticle vacua (in particular their lack of symmetry restrictions), ii) the symmetry restorations performed, and iii) the variety of the quasiparticle vacua included in the mixing. The Bogoliubov quasiparticle states used in our PGCM calculations will be referred to hereafter as \emph{intrinsic} states.

More specifically, while considering symmetry-unrestricted trial quasiparticle vacua is a necessary condition to obtain symmetry-broken intrinsic wave functions at the end of the minimization process, it is not a sufficient condition. Indeed, it is also necessary that the minimization favors a symmetry-broken solution, which in turns depends on the Hamiltonian and system considered.  
For example, the presence of deformation/pairing correlations in the intrinsic states describing a doubly-open-shell system requires a sufficient strength of the
multipole/pairing parts of the nuclear Hamiltonian at play.
There are two complementary ways of circumventing this problem and including those correlations in the intrinsic states, and they are related to points ii) and iii) above. On the one hand, the variation after projection (VAP) method~\cite{RS80a}, that minimizes the symmetry-restored energy instead of minimizing the mean-field energy, produces intrinsic states that are not, in general, eigenstates of the symmetry operators. However, the latter is costly from the computational point of view and only few implementations have been performed so far~\cite{Schmid_1987,SCHMID2004565,rodriguez-guzman_spherical_2004,ANGUIANO2001467,PhysRevC.76.014308,Maqbool_2011,PhysRevLett.114.032501,ROMERO2019177,PhysRevC.103.014312}. On the other hand, the inclusion of constraints in the HFB or VAP minimization functional allows the explicit exploration of the selected collective degrees of freedom. Moreover, this technique can be used to produce a set of intrinsic wave functions that can be subsequently projected and mixed, enriching the variational space and giving much better approximations to the exact spectrum of the nuclear Hamiltonian. 

Recently, the PGCM method demonstrated its ability to reproduce almost exact results in the $pf$-shell Ca isotopes~\cite{Bally19a}. Nevertheless, in those calculations only quadrupole deformations and neutron-neutron ($nn$) pairing were taken into account due to the absence of protons in the valence space for Ca isotopes. It is expected that having protons and neutrons simultaneously interacting will lead us to an impoverishment of the results compared with those where only one species of nucleons were present. 
In this work, we seek further understanding of the collective behavior of the nuclei, exploring how the inclusion of $pn$-pairing affects the results when variational methods are performed.

The paper is organized as follows. First, in Sec.~\ref{Theoretical_Framework}, we summarize the theoretical framework used in this work (see Ref.~\cite{TAURUS1} for more details). Then, in Sec.~\ref{Results}, we show the performance of several variational methods, starting from simple unconstrained results and finishing with full-fledged PGCM calculations. We use $^{24}$Ne (even-even), $^{25}$Ne (even-odd) and $^{24}$Na (odd-odd) as examples to illustrate the methods. Then, these calculations are extended to the full $sd$-shell  where ground and excited state energies, as well as electromagnetic transitions, are discussed. Finally, a summary is given in Sec.~\ref{Summary}.

\section{Theoretical framework}
\label{Theoretical_Framework}

In Refs.~\cite{TAURUS1,Bally19a}, we gave a detailed account of our theoretical framework. Therefore, we refer to these articles for details and only discuss here the aspects that are the most important for the present study. Also, 
from a general perspective, we mention that the exact solutions and the PGCM approximations were obtained using the codes ANTOINE~\cite{ant3} and TAURUS~\cite{TAURUS1}, respectively.

\subsection{Model space and nuclear Hamiltonian}

We consider a model space spanned by a set of spherical harmonic oscillator single-particle basis states associated with the annihilation/creation operators $\left\{ c_a ; c_a^\dagger \right\}$ and   characterized by their principal quantum number $n_a$, orbital angular momentum $l_a$, spin $s_a = 1/2$, total angular momentum $j_a$ and its third component $m_{j_a}$, and isospin $t_a = 1/2$ and its third 
component $m_{t_a}$. We use as a convention $m_{t_a} = -1/2$ for proton single-particle states and $m_{t_a} = +1/2$ for neutron single-particle states. For the sake of clarity, we use in the shorthand notation $a \equiv (n_a , l_a, s_a , j_a, m_{j_a}, t_a, m_{t_a} )$. In addition, the model space is chosen such as to be invariant under both spatial and isospin rotations.

More specifically, in this work we consider the $sd$-shell valence space defined by the orbits: $0d_{5/2}$, $1s_{1/2}$ and $0d_{3/2}$. As for the Hamiltonian, we use the well-known USD interaction~\cite{WILDENTHAL19845,USD_BROWN}. Note that because we work in a restricted model space, electromagnetic transitions and moments are calculated with the effective charge $e_{p}=1.5$ for protons and $e_{n}=0.5$ for neutrons.

\subsection{Nuclear many-body wave functions}

Within our implementation of the PGCM, the nuclear states are defined as 
\begin{equation}
|\Psi^{JMNZ\pi}_{\sigma} \rangle=\sum_{qK} f^{JMNZ\pi}_{\sigma;qK} \hat{P}^{J}_{MK} \hat{P}^{N} \hat{P}^{Z} \hat{P}^{\pi} |\Phi(q)\rangle ,
\label{GCM_wf1}
\end{equation}
where $\left\{|\Phi(q)\rangle\right\}$ are the so-called intrinsic states, assumed here to be Bogoliubov quasiparticle states (see below), 
$\hat{P}^{N(Z)}$ is the projector onto good number of neutrons (protons), 
$\hat{P}^{J}_{MK}$ is the angular momentum projection operator, and $\hat{P}^{\pi}$ is the parity projector \cite{RS80a,PhysRevC.103.024315}. To simplify the notation, we write the set of quantum numbers with the label $\Gamma \equiv (JMNZ\pi)$.
The coefficients $f^{\Gamma}_{\sigma;qK}$ are variational parameters that are obtained through the minimization of the PGCM energy that leads to the Hill-Wheeler-Griffin (HWG) equations~\cite{Griffin57a}
\begin{equation}
\sum_{q'K'} \left( \mathcal{H}^{\Gamma}_{qK,q'K'} - E^{\Gamma}_{\sigma} \mathcal{N}^{\Gamma}_{qK,q'K'} \right) f^{\Gamma}_{\sigma;q'K'} = 0 ,
\label{HWG_1}
\end{equation}
where 
\begin{subequations}
\begin{align}
\mathcal{H}^{\Gamma}_{qK,q'K'}&= \langle \Phi(q)|\hat{H}\hat{P}^{J}_{KK'}\hat{P}^{N}\hat{P}^{Z}\hat{P}^{\pi}|\Phi(q')\rangle , \\
\mathcal{N}^{\Gamma}_{qK;q'K'}&= \langle \Phi(q)|\hat{P}^{J}_{KK'}\hat{P}^{N}\hat{P}^{Z}\hat{P}^{\pi}|\Phi(q')\rangle ,
\end{align}
\end{subequations}
are the Hamiltonian and norm overlap matrices. The HWG equations are solved to obtain the energies, $E^{\Gamma}_{\sigma}$, and wave functions, $|\Psi^{JMNZ\pi}_{\sigma} \rangle$, that are used to compute other properties such as, e.g., electromagnetic transition probabilities and moments.

\subsection{Intrinsic states}
\label{sec:intrinsic}

One of the most important aspects of the PGCM method is the choice of the intrinsic wave functions $\left\{|\Phi(q)\rangle\right\}$.
In the present case, all of them are Bogoliubov quasiparticle wave functions, i.e., they are vacua for a set of quasiparticle operators $\left\{ \beta_a (q) ; \beta^{\dagger}_a (q) \right\}$
defined through unitary linear Bogoliubov transformations
\begin{equation}
\beta^{\dagger}_{b}(q)=\sum_{a}U_{ab}(q)c^{\dagger}_{a}+V_{ab}(q)c_{a} .
\label{HFB_trans}
\end{equation} 
   
The matrix elements of $U(q)$ and $V(q)$ are variational parameters that are determined by minimizing the energy of the system. Here, we minimize either the HFB energy or the variation after particle-number projection (VAPNP) energy with constraints on certain parameters labelled generically as $q$ (see below for details). 

Due to computational reasons, it is customary to impose some symmetry restrictions on the Bogoliubov vacua that are self-consistently conserved (up to numerical accuracy) throughout the minimization process if no symmetry-breaking constraint is used. Examples of such \emph{intrinsic symmetries} are the axial symmetry, parity, or prohibiting proton/neutron mixing. In this work, we only impose that the matrices $U(q)$ and $V(q)$ are real in the definition of the Bogoliubov transformations. In particular, the transformations include $pn$-mixing that allows the appearance of $pn$-pairing terms. Nevertheless, because we work in the $sd$-shell valence space that contains only positive-parity single-particle states, our intrinsic states automatically inherit a good parity and no parity projection is required when building the PGCM wave functions.

While a general trial wave function is always better from a pure variational point of view, the use of more restricted seed wave functions can be of great interest to study the relevance of the broken symmetries in the description of nuclear spectra and other observables. In particular, in order to understand better the importance of mixing protons and neutrons, we use three types of initial wave functions in the HFB and VAPNP calculations:
\begin{itemize}
    \item real axially symmetric Bogoliubov vacua without $pn$-mixing, labelled as \textit{axial $pn$-no}.
    \item real general Bogoliubov vacua without $pn$-mixing, labelled as \textit{general $pn$-no}.
    \item real general Bogoliubov vacua with $pn$-mixing, labelled as \textit{general $pn$-yes}.
\end{itemize}

A quantitative way of checking that the HFB or VAPNP intrinsic wave functions incorporate $pn$-pairing correlations is the evaluation of the mean-field $pp$-, $nn$- and $pn$-pairing energies defined as~\cite{RS80a}
\begin{equation}
E_{\mathrm{pair}}=E^{pp}_{\mathrm{pair}}+E^{nn}_{\mathrm{pair}}+2E^{pn}_{\mathrm{pair}}
\end{equation}
where 
\begin{equation}
E^{\tau\tau'}_{\mathrm{pair}}=\frac{1}{2}\sum_{a_{\tau}b_{\tau'}}\Delta_{a_{\tau}b_{\tau'}}\kappa^{*}_{a_{\tau}b_{\tau'}}
\end{equation}
and the isospin ($\tau=p/n$) dependencies of the pairing density and pairing field are written explicitly as
\begin{eqnarray}
\kappa_{a_{\tau}b_{\tau'}}&=&\langle\Phi(q)|c_{b_{\tau'}}c_{a_{\tau}}|\Phi (q)\rangle ,\\
\Delta_{a_{\tau}b_{\tau'}}&=&\frac{1}{2}\sum_{c_{\tau''}d_{\tau'''}}\bar{v}_{a_{\tau}b_{\tau'}c_{\tau''}d_{\tau'''}}\kappa_{c_{\tau''}d_{\tau'''}} .
\end{eqnarray}
Here, $\bar{v}_{abcd}$ are the antisymmetrized two-body matrix elements of the nuclear interaction in the working basis. It is important to point out that $E^{pn}_{\mathrm{pair}} = 0$ for Bogoliubov transformations that do not include $pn$-mixing, i.e., with transformations of the form: $U_{a_{\tau}b_{\tau'}}=U_{a_{\tau}b_{\tau}}\delta_{\tau\tau'}$ and $V_{a_{\tau}b_{\tau'}}=V_{a_{\tau}b_{\tau}}\delta_{\tau\tau'}$.

All Bogoliubov vacua, even the most general ones, always conserve the number-parity symmetry associated to a subgroup of nucleon gauge rotations and as such can be characterized by their nucleon number-parity quantum number \cite{PhysRevC.103.024315}. More specifically, there exist Bogoliubov vacua with either an even or odd number parity. The former can be used to described nuclear systems made of an even number of particles wheras the latter describe systems made of an odd number of particles. If the Bogoliubov transformations do not allow for $pn$-mixing, the nucleon number parity can be factorized as the product of the number parity for protons times the one for neutrons.
When describing even-odd or odd-odd systems, and assuming that one starts from a Bogoliubov vacua with an even nucleon number parity, one has to be careful about the quasiparticle one decides to block to obtain a wave function with the correct structure. These issues will be discussed later on.

Finally, we consider the selection of the collective coordinates, $q$. These are imposed during the minimization by using constraints on the expectation values of the operators $\hat{Q}$ associated to the collective coordinates., i.e.\ $\langle \Phi(q)|\hat{Q}|\Phi(q)\rangle=q$, with the Lagrange multipliers $\lambda_Q$.
Hence, the HFB and VAPNP minimization functionals are defined respectively as
\begin{align}
E'_{\mathrm{HFB}}\left[|\Phi(q)\rangle\right] = &\langle\Phi(q)|\hat{H}-\lambda_{N}\hat{N}-\lambda_{Z}\hat{Z}-\lambda_{q}\hat{Q}|\Phi(q)\rangle,\nonumber\\ \\
E'_{\mathrm{VAPNP}}\left[|\Phi(q)\rangle\right] = &\frac{\langle\Phi(q)|\hat{H}\hat{P}^{N}\hat{P}^{Z}|\Phi(q)\rangle}{\langle\Phi(q)|\hat{P}^{N}\hat{P}^{Z}|\Phi(q)\rangle}\nonumber \\
& - \langle\Phi(q)|\lambda_{q}\hat{Q}|\Phi(q)\rangle.
\end{align}
Here, $\lambda_{N(Z)}$ are also Lagrange multipliers that guarantee the condition $\langle \Phi(q)|\hat{N}(\hat{Z})|\Phi(q)\rangle=N(Z)$ at the HFB level. 
In the end, the variational problem to solve at the HFB(VAPNP) level is given by 
\begin{equation}
    \delta[E'_{\mathrm{HFB(VAPNP)}}] = 0 .
\end{equation}

In this work, we explore both pairing and quadrupole degrees of freedom since they are the most important terms in the multipole decomposition of realistic nuclear interactions~\cite{PhysRevC.54.1641}. 
The former can be explored in a general way with operators that couple pairs of nucleons in a given orbit, $\breve{a} \equiv  (n_a, l_a , j_a , s_a , t_a)$, to a good total isospin and total angular momentum of the pair, $J_p T_p$~\cite{talmi,PhysRevC.90.031301}
\begin{equation}
\begin{split}
\left[\hat{P}^{\dagger}\right]^{J_p T_p}_{M_{J_p}M_{T_p}} &= \sum_{\breve{a}} \left[\hat{P}^{\dagger}_{\breve{a}}\right]^{J_p T_p}_{M_{J_p}M_{T_p}} \\
&= \frac{1}{\sqrt{2}}\sum_{\breve{a}}\sqrt{2j_{a}+1}\left[c^{\dagger}_{\breve{a}}c^{\dagger}_{\breve{a}}\right]^{J_p T_p}_{M_{J_p}M_{T_p}}
\end{split}
\end{equation}
where the creation operators are $J_p T_p$-coupled according to 
\begin{align}
\left[c^{\dagger}_{\breve{a}}c^{\dagger}_{\breve{b}}\right]^{J_p T_p}_{M_{J_p}M_{T_p}} 
&= \frac{\sqrt{1-\delta_{\breve{a} \breve{b}}(-1)^{J_p + T_p}}}{1+\delta_{\breve{a}\breve{b}}}\sum_{\substack{m_{j_{a}}m_{j_{b}} \\ m_{t_{a}}m_{t_{b}}}} c^{\dagger}_{a}c^{\dagger}_{b} \\
& \times \langle j_{a}m_{j_{a}}j_{b}m_{j_{b}}|J_p M_{J_p}\rangle \langle \tfrac{1}{2}m_{t_{a}}\tfrac{1}{2}m_{t_{b}}|T_p M_{T_p}\rangle . \nonumber 
\end{align}
The isovector channel ($J_p=0$, $T_p=1$) can be used to explore the usual $pp$-, $nn$-pairing correlations as well as a part of the $pn$ pairing. On the other hand, the isoscalar channel ($J_p=1$, $T_p=0$) is purely associated with $pn$ pairing.
More specifically, we define the parameters $\delta^{J_p T_p}_{M_{J_p}M_{T_p}}$ as
\begin{equation}
\delta^{J_p T_p}_{M_{J_p}M_{T_p}}= \frac12 \langle \Phi(q) | \left[\hat{P}\right]^{J_p T_p}_{M_{J_p}M_{T_p}} +
\left[\hat{P}^{\dagger}\right]^{J_p T_p}_{M_{J_p}M_{T_p}} | \Phi(q) \rangle
\label{pair_op}
\end{equation}
that measures the amount of a certain type of pairing correlations in the intrinsic wave function.

Concerning quadrupole deformations, they can be explored by imposing constraints on the average values of the mass quadrupole operators
\begin{subequations}
\begin{align}
\hat{Q}_{2\mu} &= r^{2} Y_{2\mu}(\theta,\varphi) , \\
q_{20} &= \langle\Phi(q)|\hat{Q}_{20}|\Phi(q)\rangle ,\\
q_{21} &= \frac12 \langle\Phi(q)|\hat{Q}_{21}-\hat{Q}_{2-1}|\Phi(q)\rangle , \\
q_{22} &= \frac12 \langle\Phi(q)|\hat{Q}_{22}+\hat{Q}_{2-2}|\Phi(q)\rangle ,
\end{align}
where $Y_{2\mu}(\theta,\varphi)$ is a spherical harmonics of degree 2 and order $\mu$.
In the present calculations, except told otherwise, we always set $q_{21}=0$. In such a case, we can also define the triaxial parameters $\overline{\beta}_{2}$~\footnote{We use the notation $\overline{\beta}_{2}$ to distinguish the deformations calculated in a valence-space, but without using effective charges in the definition of the quadrupole operators, from the regular $\beta_{2}$ values used in no-core calculations where all the nucleons contribute to the deformation of the nucleus.} and $\gamma$ as
\begin{align}
\bar{\beta_{2}} &= C \sqrt{q_{20}^2 + 2 q_{22}^2} , \\
\gamma &= \arctan\left(\frac{\sqrt{2} q_{22}}{q_{20}} \right) , 
\end{align}
\end{subequations}
where $C=\frac{4\pi}{3r_{0}^{2}A^{5/3}}$, $r_{0}=1.2$ fm and $A$ is the total mass number (including core and valence space particles). 

\section{Results}
\label{Results}
\subsection{Unconstrained calculations}
\label{unconstrained_calc}
We start by analyzing the absolute HFB and VAPNP minima obtained with the different types of trial Bogoliubov vacua considered in Sec.~\ref{sec:intrinsic}. As an illustrative case, we will study first the unconstrained solutions 
obtained for three nuclei: the even-even (e-e) nucleus $^{24}$Ne, the even-odd (e-o) nucleus $^{25}$Ne and the odd-odd (o-o) nucleus $^{24}$Na. Then, we will extend the calculations to the whole $sd$-shell.
\subsubsection{$^{24}$Ne (e-e) case}
\begin{figure}[t]
\begin{center}
\includegraphics[width=\columnwidth]{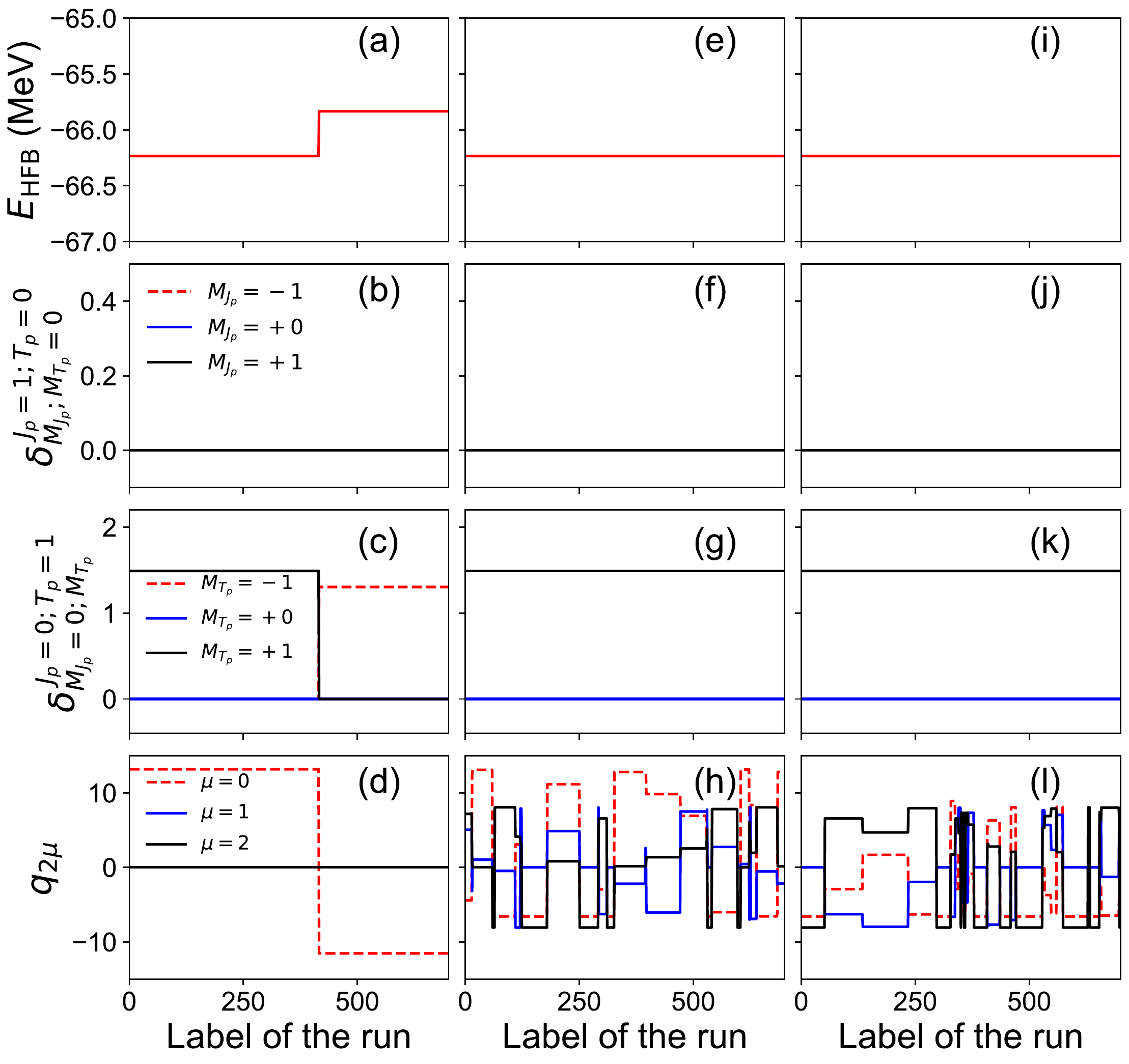}
\end{center}
\caption{(color online) Total HFB energy (first row), isoscalar (second row) and isovector (third row) pairing amplitudes, and quadrupole moments (fourth row) obtained in 650 solutions of the unconstrained HFB equations with (a)-(d) random axial $pn$-no seeds, (e)-(h) random general $pn$-no seeds and (i)-(l) random general $pn$-yes seeds. The nucleus considered is $^{24}$Ne ($Z=2$ and $N=6$ in the valence space) and was calculated using the USD interaction.}
\label{24Ne_minimum_hfb}
\end{figure}

\begin{figure}[t]
\begin{center}
\includegraphics[width=\columnwidth]{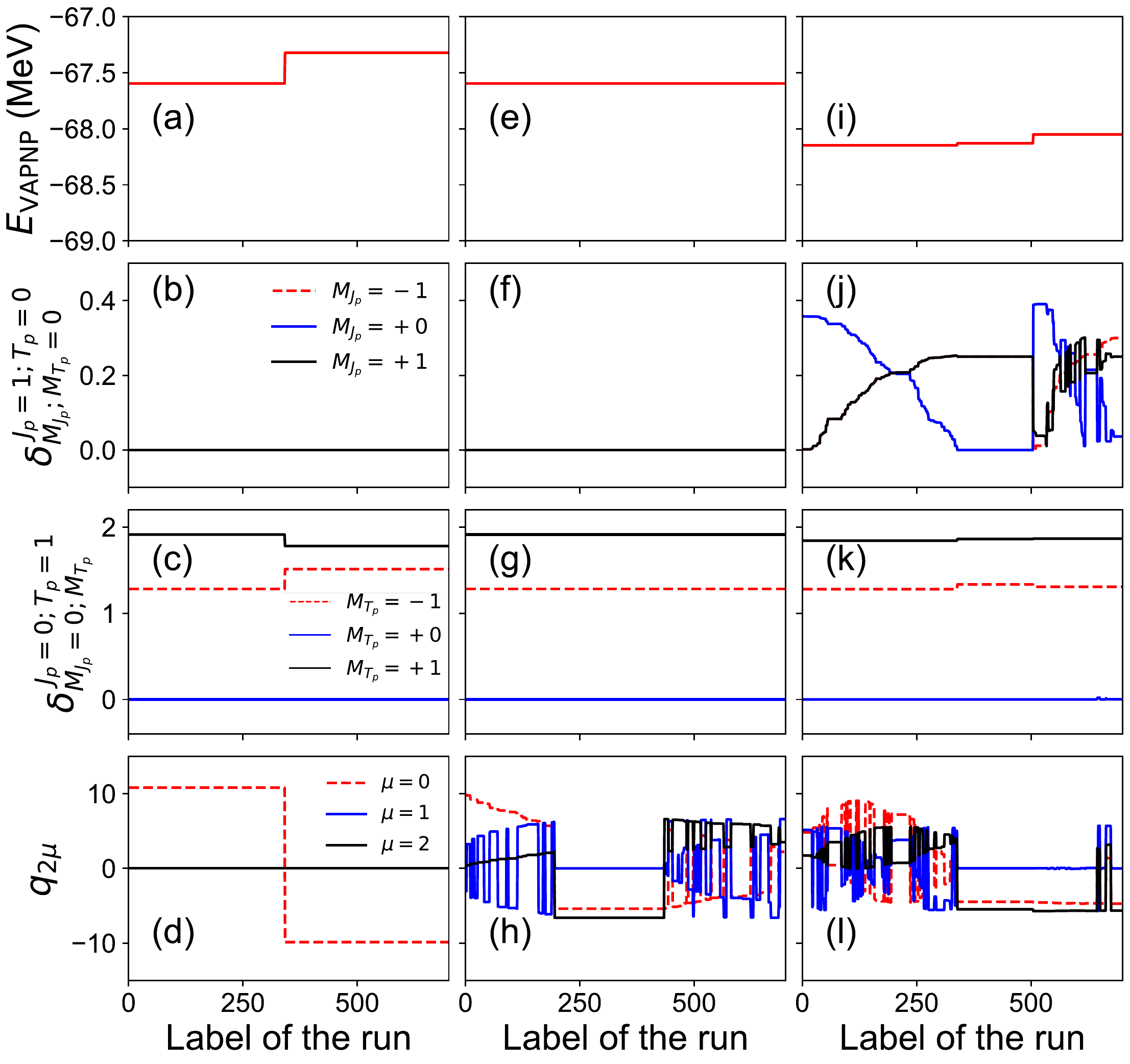}
\end{center}
\caption{(color online) Same as Fig.~\ref{24Ne_minimum_hfb} but for the VAPNP method.}
\label{24Ne_minimum_VAPNP}
\end{figure}

Since the HFB and VAPNP equations are solved using a first-order gradient method, the algorithm can lead to several minima (if there exist) depending on the choice of the seed wave function and on the energy surface of the system. A way to bypass this problem and ensure that the absolute energy minimum is reached, is to repeat the calculation several times starting from different random seed wave functions. The results obtained by performing 650 runs of the HFB and VAPNP calculations are shown in Figs.~\ref{24Ne_minimum_hfb} and~\ref{24Ne_minimum_VAPNP}, respectively.
In addition to the HFB/VAPNP energies, we show the expectation values of the pairing and quadrupole operators that are useful to interpret the results. In particular, for these calculations, $q_{21}$ is not constrained to zero. In the HFB case,  
we observe that the absolute minimum does not depend on the choice of structure for the seed wave function, i.e., the axial and the two general seeds lead to the same variational energy. 
In fact, the solutions do not display neither $pn$-pairing (isoscalar and isovector) nor $pp$-pairing, even when the variational space
 would allow for the inclusion of some (or all) of these correlations.
Only $nn$-pairing correlations ($\delta^{J_p=0;T_p=1}_{M_{J_p}=0;M_{T_p}=+1}$) are different from zero.
 In addition, we observe that when using an axial seed, one obtain both the absolute prolate solution ($q_{20}>0$) and also 
another local minimum with an oblate deformation ($q_{20}<0$). Obviously, because of the axial symmetry (along the $z$ axis) we have $q_{2\mu}=0$ for $\mu=1,2$ in both cases. By contrast, using the general $pn$-no and $pn$-yes seeds is more efficient in finding the global minimum although the quadrupole parameters $q_{2\mu}$ oscillate [Fig.~\ref{24Ne_minimum_hfb}(h)-(l)]. But this oscillation only reflects that the energy does not depend on the orientation of the deformed nucleus with respect to the coordinate system.

Considering now the VAPNP calculations, the minimizations without the possibility of exploring the proton-neutron mixing [Fig.~\ref{24Ne_minimum_VAPNP}(a)-(h)] give again the same lowest energy with the same amount of pairing correlations. Contrary to the HFB case, here both $pp$- and $nn$-pairing correlations are obtained thanks to the superior variational scheme that minimizes the particle-number-projected energy. Using axial seeds, the minimum corresponds to a prolate state and a secondary minimum appears at an oblate configuration. The prolate deformed state is also obtained with the general $pn$-no solutions but the quadrupole parameters are rearranged due to the arbitrary orientation of the wave function previously discussed. 
The main differences with the HFB calculations are observed whenever a general seed that allows for $pn$-mixing is used [Fig.~\ref{24Ne_minimum_VAPNP}(i)-(l)]. First, these calculations provide with the absolute energy minimum of all calculations. Second,
in this case not only $pp$ and $nn$ isovector but also $pn$ isoscalar pairing correlations are present in the intrinsic wave function. 
We also see two additional energy minima very close to the absolute one. Finally, both the quadrupole deformation ($q_{2\mu}$) and isoscalar pairing $(J_p=1, T_p=0, M_{J_p})$ parameters change due to the arbitrary orientation of the state, but again the total energy remains constant.  

\begin{figure}[t]
\begin{center}
\includegraphics[width=\columnwidth]{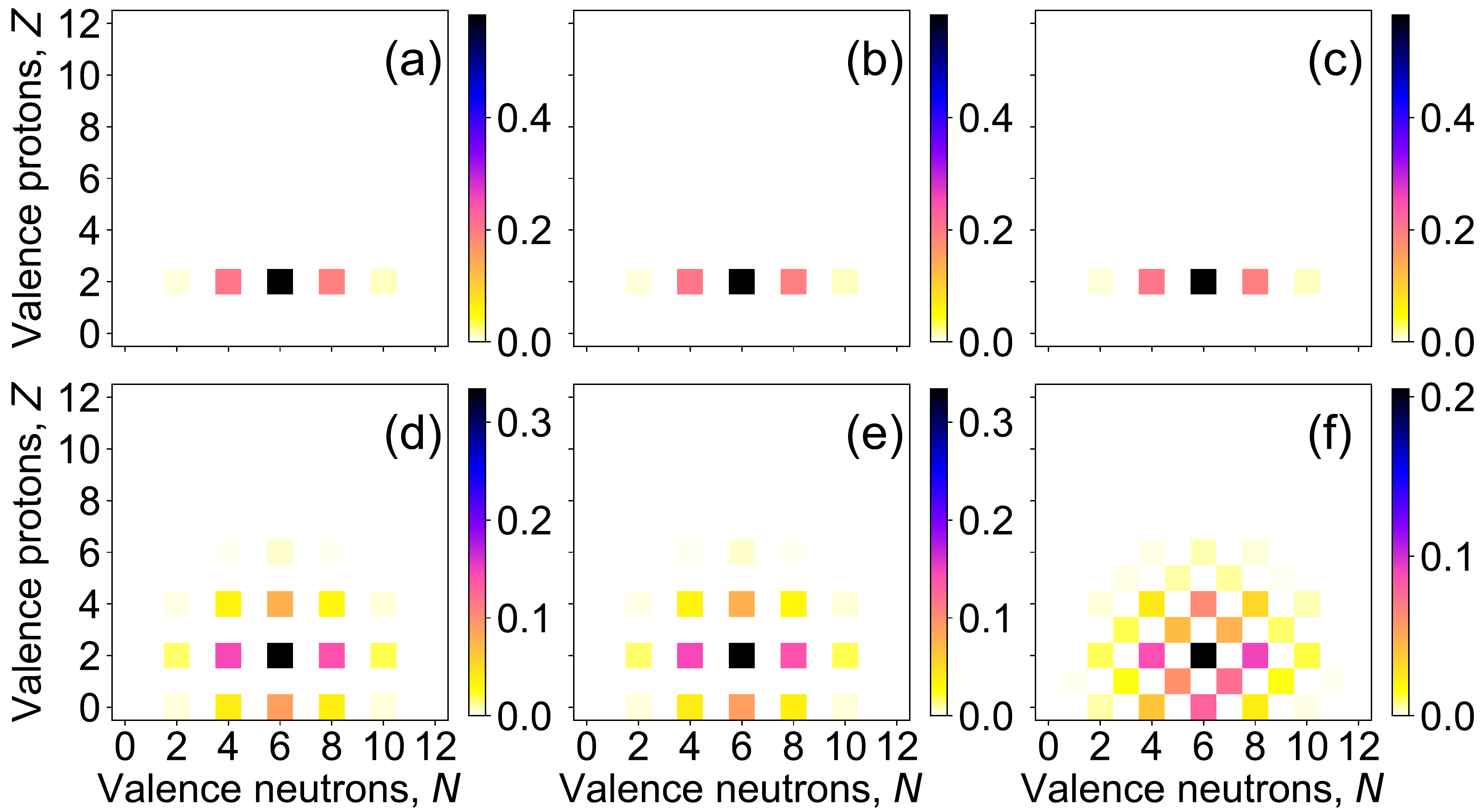}
\end{center}
\caption{(color online) Distribution in eigenstates of the proton and neutron number operators of the intrinsic wave functions ($\langle\Phi|\hat{P}^{Z}\hat{P}^{N}|\Phi\rangle$) obtained from minimizing HFB/VAPNP energy functionals with different seed wave functions: (a) HFB axial $pn$-no, (b) HFB general $pn$-no, (c) HFB general $pn$-yes, (d) VAPNP axial $pn$-no, (e) VAPNP general $pn$-no, and, (f) VAPNP general $pn$-yes. The nucleus considered is $^{24}$Ne ($Z=2$ and $N=6$ in the valence space) and was calculated using the USD Hamiltonian.}
\label{24Ne_minimum_nz_dist}
\end{figure}

We can now analyze in more detail the consequences of having pairing correlations in the wave functions by decomposing them in terms of eigenstates of the proton and neutron number operators. This decomposition is obtained by computing the particle number projected norm overlap, $\langle\Phi|\hat{P}^{Z}\hat{P}^{N}|\Phi\rangle$, for all possible values of ($Z,N$) \cite{PhysRevC.103.024315}. Furthermore, there exist the sum rule $\sum_{ZN}\langle\Phi|\hat{P}^{Z}\hat{P}^{N}|\Phi\rangle=1$ due to the normalization of the intrinsic wave function. In Fig.~\ref{24Ne_minimum_nz_dist}, we show such distributions. In all cases, the maximum of the distribution is obtained at the values of the targeted nucleus $^{24}$Ne [$(Z_{0},N_{0})=(2,6)$ in the valence space]. The rest of the distribution depends on the functional that is minimized. For this nucleus we see that the HFB results [Fig.~\ref{24Ne_minimum_nz_dist}(a)-(c)] are independent of the type of seed used in the calculation. Due to the absence of $pp$-pairing correlations, only eigenstates with $Z=2$ are present in these cases. In addition, eigenstates with $6\pm2$ and $6\pm4$ neutrons are also found but with significantly smaller weights. Concerning the VAPNP results [Fig.~\ref{24Ne_minimum_nz_dist}(d)-(f)], we see a wider distributions of components, which can be related to the fact that both $pp$- and $nn$-pairing correlations are present. The distributions obtained with the seeds that do not mix protons and neutrons are the same because the absolute minimum is axially symmetric. Moreover, since there are no $pn$-pairing correlations in this case, only eigenstates with ($Z=$ even, $N=$ even) are obtained. The weights decrease almost like a binomial distribution around the maximum value at $(Z,N)=(2,6)$, which can be justified theoretically \cite{Flocard97a} and was also recently exemplified in Ref.~\cite{PhysRevC.103.024315}.
 More interestingly, the presence of $pn$-pairing correlations in the VAPNP solution obtained from the seed general $pn$-yes [Fig.~\ref{24Ne_minimum_nz_dist}(f)] produce a distribution where both ($Z=$ even, $N=$ even) and ($Z=$ odd, $N=$ odd) eigenstates are found. Although the contributions of the latter are slightly smaller than the one of the e-e neighbors, they are not negligible. As we will see in subsequent sections, this fact opens the possibility of studying odd-odd nuclei without blocking two quasiparticles explicitly. 
\subsubsection{$^{25}$Ne (e-o) case}

As mentioned in Sec.~\ref{Theoretical_Framework}, even-odd and odd-even nuclei must be described by intrinsic wave functions that have an odd number parity. Hence, this case introduces another factor apart from the choice of the energy functional and seed wave function, i.e., the selection of the initial blocked state. We first discuss the influence of the initial blocked state on the final HFB/VAPNP energies for the different seeds. These initial states are defined as $|\Phi_{0,a}\rangle=\beta^{\dagger}_{0,a}|\Phi_{0}\rangle$, where $|\Phi_{0}\rangle$ is an even-even seed of the same type as those used for $^{24}$Ne (axial and general $pn$-no, general $pn$-yes), and $\beta^{\dagger}_{0,a}$ is a quasiparticle creation operator~\cite{PhysRevLett.113.162501,bally1,PhysRevC.98.044317}. For $^{25}$Ne in the $sd$-shell, we have 12 possible choices for the initial neutron blocked state.
In addition, when starting from an axially symmetric $|\Phi_{0}\rangle$, the even-odd wave function can be characterized by its angular-momentum component along the $z$ axis, i.e., $\hat{J}_{z}|\Phi_{a}\rangle=K_{a}|\Phi_{a}\rangle$. 
As in the e-e case (see Figs.~\ref{24Ne_minimum_hfb} and \ref{24Ne_minimum_VAPNP}), we performed a series of HFB/VAPNP calculations (650 runs) with different seeds for the 12 possibilities of the initial blocked state. The results (not shown) are similar to those obtained for $^{24}$Ne and we will not discuss them further.
\begin{figure}[t]
\begin{center}
\includegraphics[width=\columnwidth]{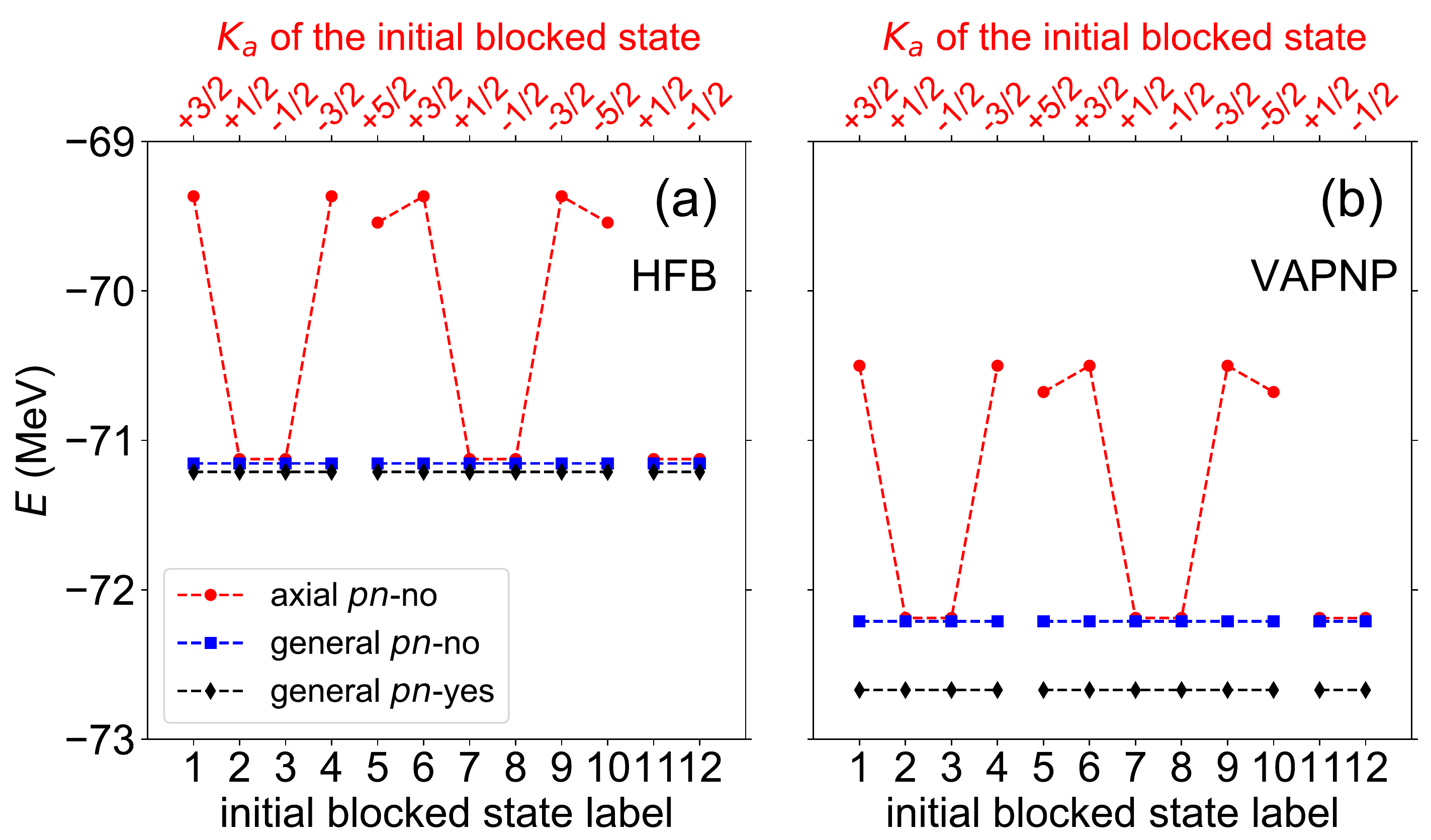}
\end{center}
\caption{(color online) (a) HFB and (b) VAPNP energies calculated with axial $pn$-no (red bullets), general $pn$-no (blue squares) and general $pn$-yes (black diamonds) seeds as a function of the label  of the initial blocked neutron level in the $sd$-shell (from left to right: $0d_{3/2}$, $0d_{5/2}$, $1s_{1/2}$ orbits). The nucleus considered is $^{25}$Ne ($Z=2$ and $N=7$ in the valence space) and was calculated using the USD Hamiltonian.}
\label{25Ne_minimum_blocking_h}
\end{figure}
Taking only the absolute minima in each case, we represent in Fig.~\ref{25Ne_minimum_blocking_h} those energies as a function of the initial blocked state. Here, we notice that the VAPNP energy is below the HFB energy when considering the same type of seed. We also see that the solution based on general $pn$-yes seeds is below the solution based on general $pn$-no seeds, the latter being itself below the energy obtained with the more restricted axial $pn$-no seeds. This is because of the extra $pn$-pairing correlations included in the wave functions thanks to the fact that $pn$-mixing are allowed in the Bogoliubov transformations. More interestingly, we observe that only in the axial case there is a strong dependence of the energy on the initial blocked state. In such cases, not only the initial but also the final states have a good $K_a$ quantum number and we find the same energies for the same initial $|K_{a}|$-value. In fact, for this particular nucleus, the lowest energy within the axial approximation is found at $|K_{{a}}|=1/2$, which is the expected total angular momentum for the ground state of $^{25}$Ne in the naive shell model picture. Therefore, using general seeds simplifies the calculations because we do not have to worry (in most of the cases) about the choice of the initial blocked level.  

\begin{figure}[t]
\begin{center}
\includegraphics[width=\columnwidth]{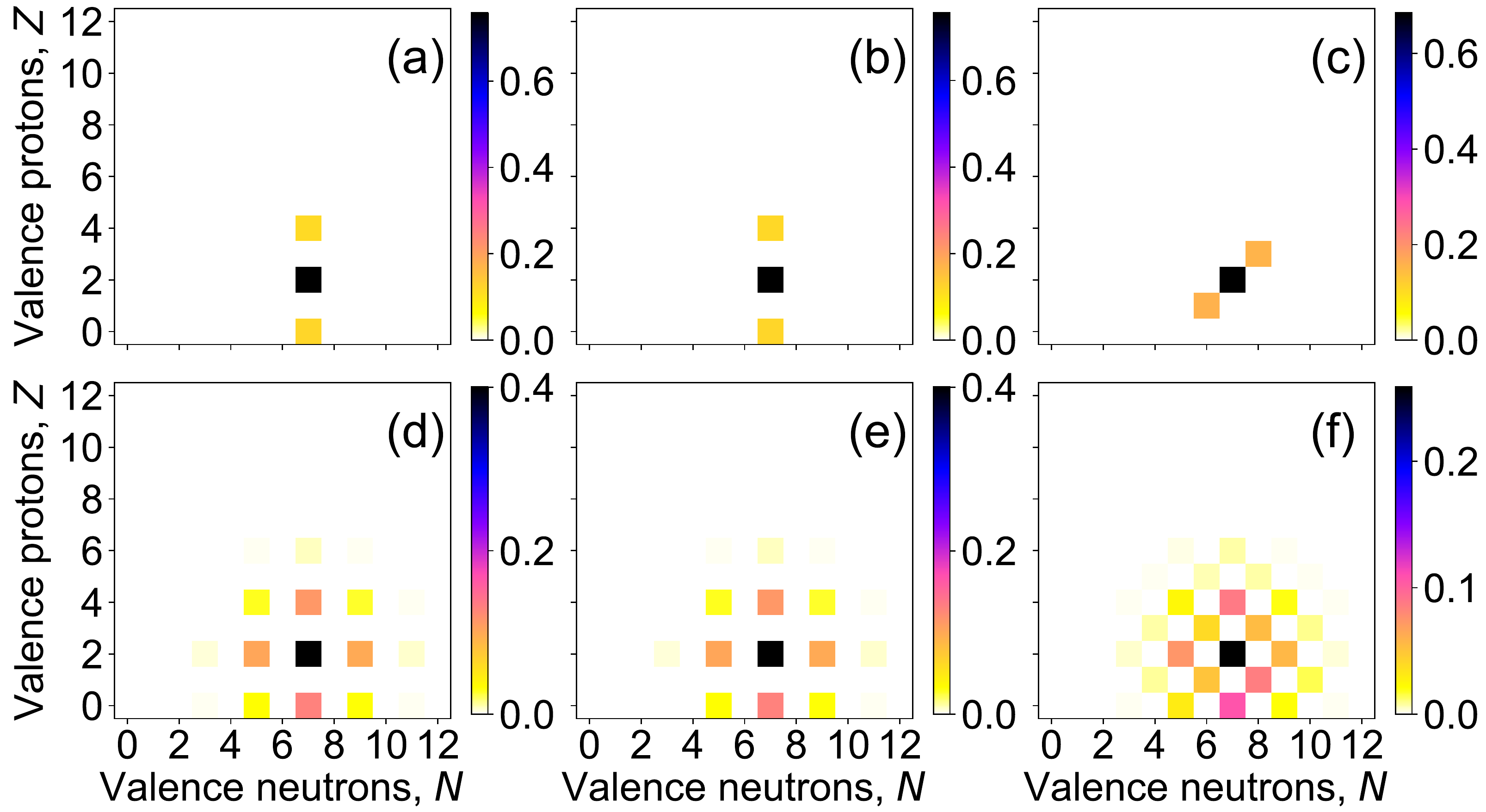}
\end{center}
\caption{(color online) Same as Fig.~\ref{24Ne_minimum_nz_dist} but for the $^{25}$Ne nucleus ($Z=2$ and $N=7$ in the valence space).}
\label{25Ne_minimum_nz_dist}
\end{figure}
We now analyze the decomposition of the HFB/VAPNP unconstrained solutions in terms of eigenstates of the particle number operators. In Fig.~\ref{25Ne_minimum_nz_dist}, we show the results obtained for the different seeds. In general, only eigenstates with $N+Z=$ odd contribute to the intrinsic wave function because the number parity in this case is odd. Moreover, the largest contribution corresponds to the targeted nucleus $^{25}$Ne ($Z_{0}=2$ and $N_{0}=7$ in the valence space). In the the decomposition of the HFB solutions based on seeds without $pn$-mixing [Fig.~\ref{25Ne_minimum_nz_dist}(a)-(b)] we find only eigenstates with $N=7$ since the blocking of a neutron state results in vanishing neutron pairing correlations at the mean-field level. We also see small components with $Z=Z_{0}\pm2$. By contrast, if the $pn$-mixing is allowed, the HFB solution acquires a small isoscalar pairing correlation, but no isovector one. This internal configuration produces the distribution shown in [Fig.~\ref{25Ne_minimum_nz_dist}(c)]. Similarly to the results shown in $^{24}$Ne, richer structures are obtained when using the VAPNP method because pairing correlations are better explored. Using seeds without $pn$-mixing, the wave functions are distributed among $Z=Z_{0} \pm 2m$ and $N=N_{0} \pm 2n$ eigenstates, $m$ and $n$ being integers. Additionally, the inclusion of $pn$-mixing in the seed wave function leads to a solution with $pp$-, $nn$- and $pn$- (both $T_p=0$ and 1) pairing correlations at the same time. This is nicely reflected in Fig.~\ref{25Ne_minimum_nz_dist}(f) where the wave function can be decomposed into a large variety of e-o and o-e nuclei around $^{25}$Ne, with their weights decreasing as one moves away from the physical components $Z_0$ and $N_0$. 
\subsubsection{$^{24}$Na (o-o) case}
\begin{figure}[t]
\begin{center}
\includegraphics[width=\columnwidth]{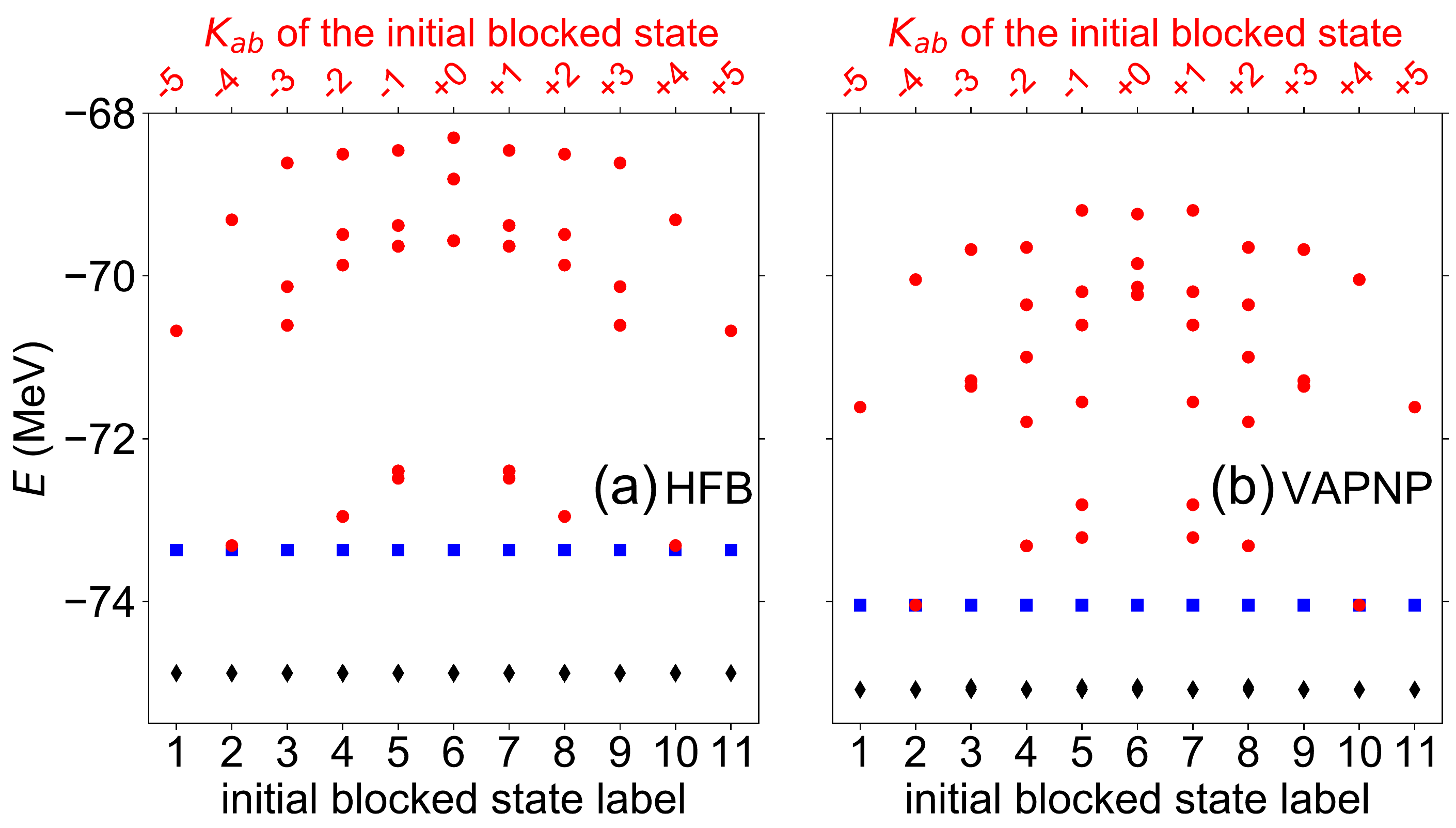}
\end{center}
\caption{(color online) (a) HFB and (b) VAPNP energies calculated with axial $pn$-no (red bullets), general $pn$-no (blue squares) and general $pn$-yes (black diamonds) seeds as a function of the label of the initial blocking levels (protons and neutrons) in the $sd$-shell. The nucleus considered is $^{24}$Na ($Z=3$ and $N=5$ in the valence space) and was calculated with the USD Hamiltonian.}
\label{24Na_minimum_blocking_h}
\end{figure}

Finally, we describe the calculations of the most involved case for HFB-based theories, i.e., the case of a nucleus with an odd number of both protons and neutrons. As in the two previous sections, we perform HFB/VAPNP calculations with different seeds and all possible blocking structures. The latter can be done because of the small dimensionality of the $sd$-shell ($12\times12=144$ possibilities). Hence, the first step is the blocking of one proton and one neutron quasiparticle to obtain a formally two-quasiparticle intrinsic wave function, $|\Phi_{0,a_{p}b_{n}}\rangle=\beta^{\dagger}_{0,a_{p}}\beta^{\dagger}_{0,b_{n}}|\Phi_{0}\rangle$, with a global even number parity. If protons and neutrons are kept separate, the state has an odd number parity for protons and neutrons separately. Moreover, if one starts from an axially symmetric $|\Phi_{0}\rangle$, one can label the blocking structure by its initial $K_{ab}=K_{a_{p}}+K_{b_{n}}$ value. Of course, the same value of $K_{ab}$ can be obtained in several ways, e.g., $K=+3=1/2_{p}+5/2_{n}=5/2_{p}+1/2_{n}=3/2_{p}+3/2_{n}$.
As in the previous e-e and e-o cases, we performed a series of calculations with random seeds (300 runs) for each combination of: a) HFB/VAPNP, b) type of seed, and, c) blocking structure, to determine better the absolute minimum for each set. Similar results to those discussed for the e-e case have been obtained (not shown). We show in Fig.~\ref{24Na_minimum_blocking_h}(a)-(b) the HFB and VAPNP minimum energies as a function of the different blocking combinations and for different structures of the seed wave functions. For the sake of simplicity in the labelling of the $x$-axis, we have grouped the 144 possibilities as if the axial symmetry were preserved. That means that the label $i$ corresponds to the same $(a_{p};b_{n})$ initial blocking columns in the $(U,V)$ matrices that produce, in the axial case, an state with $K_{ab}$. 

The general observations discussed above for the e-e and e-o cases also hold in the o-o case, i.e., VAPNP energies are lower than HFB energies, the general seeds with $pn$-mixing provide the lowest energies whereas the axially symmetric ones provide the highest energies. Moreover, the results are once again independent of the choice of the blocking structure in the case of general seeds whereas they strongly dependent on the value of the initial $K_{ab}$ in the axial case. More precisely, in the latter case the results are symmetric with respect to $\pm K_{ab}$ and, similarly to the e-o case, they only depend upon the $K_{ab}=K_{a_{p}}+K_{b_{n}}$ combination.
Finally, it is also interesting to note that the lowest energy obtained with axial blocked seeds is found at $|K|=4$ which is precisely the total angular momentum of the exact (and experimental) ground state for $^{24}$Na. 

\begin{figure}[t]
\begin{center}
\includegraphics[width=\columnwidth]{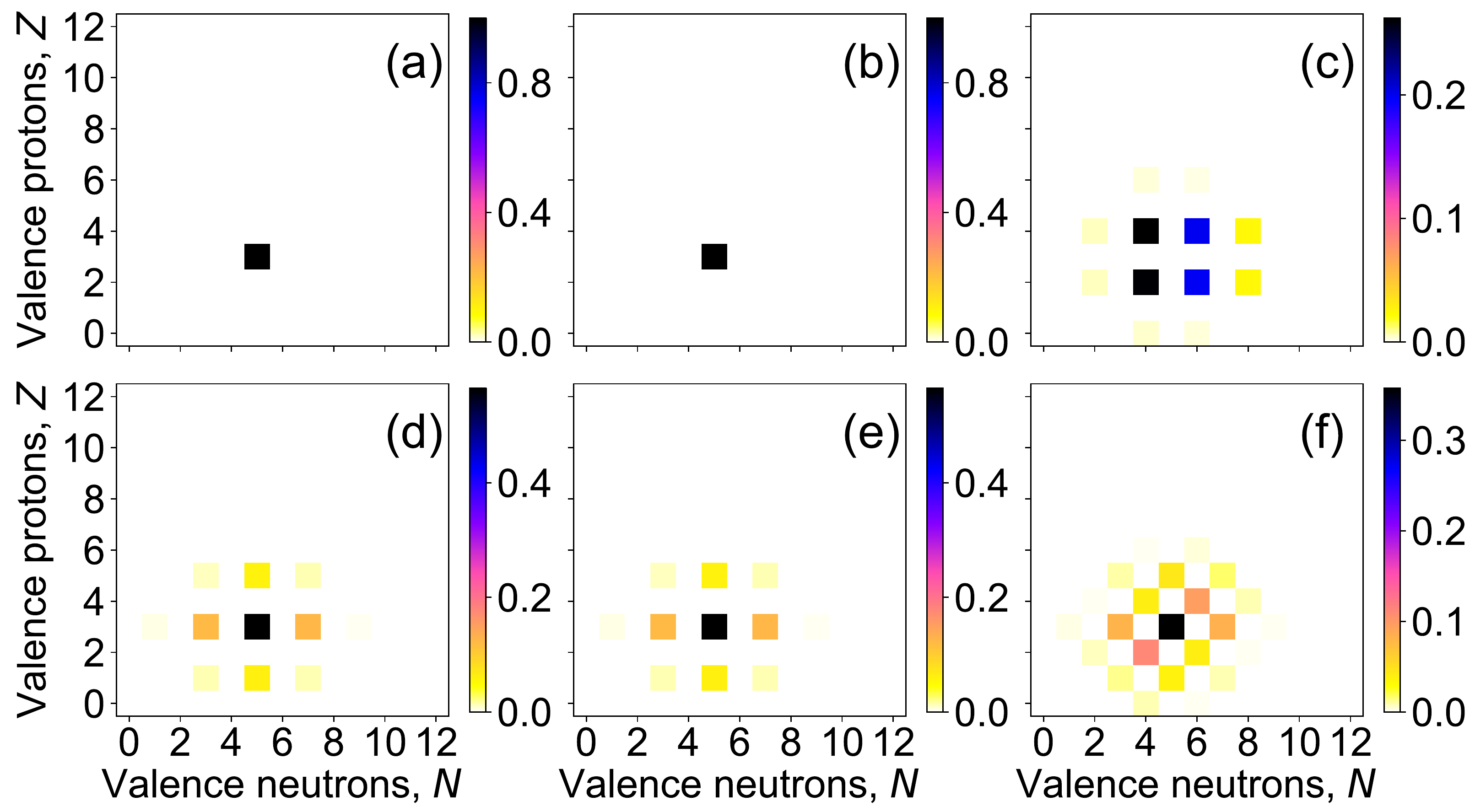}
\end{center}
\caption{(color online) Same as Fig.~\ref{24Ne_minimum_nz_dist} but for the $^{24}$Na ($Z=3$ and $N=5$ in the valence space).}
\label{24Na_minimum_nz_dist}
\end{figure}

For the sake of completeness, we represent in Fig.~\ref{24Na_minimum_nz_dist} the distribution in terms of eigenstates of the $\hat{N}$ and $\hat{Z}$ operators for the different minima discussed above. The main differences with the previous analyses are observed in the HFB method. First of all, seeds without $pn$-mixing are not able to spontaneously break the particle-number invariance and produce solutions without any kind of pairing correlations. More interestingly, the HFB solution obtained with general $pn$-yes seeds is in fact a fake odd-odd vacuum. Indeed, it does not contain any eigenstates with an odd number of protons or neutrons but the largest contributions correspond to even-even nuclei with $Z=Z_{0}\pm1$ and $N=N_{0}\pm1$ (with $(Z_{0},N_{0})=(3,5)$ for $^{24}$Na).
It is important to remark that only the nucleon number parity is enforced by the Bogoliubov transformations in this case.
Therefore, the HFB minimization produces a wave function with global even number parity and fully paired (without fully occupied single-particle states). In this sense, the HFB method with general seeds and with $pn$-mixing is useless to describe o-o nuclei. Additional constraints on the pairing content of the intrinsic states imposed during the minimization procedure could help to bypass this problem. 

Fortunately, the VAPNP method is able to describe o-o nuclei both without and with $pn$-mixing. In the first case, odd number parity both for protons and neutrons separately are safely defined. Here, $pp$ and $nn$ correlations are present [see Figs.~\ref{24Na_minimum_nz_dist}(d)-(e)] and, apart from the largest contribution at $(Z_{0},N_{0})$, we also see some components in $(Z_{0}\pm2,N_{0}\pm2)$. Finally, we obtain pairing correlations of all kind (isoscalar and isovector) when considering the most general transformation. As in the e-o case, this is reflected in the appearance of all possible types of eigenstates in the intrinsic wave function around the maximum at $(Z_{0},N_{0})$.

\subsubsection{Systematic calculations of the ground-state energies}

\begin{figure}[t]
\begin{center}
\includegraphics[width=\columnwidth]{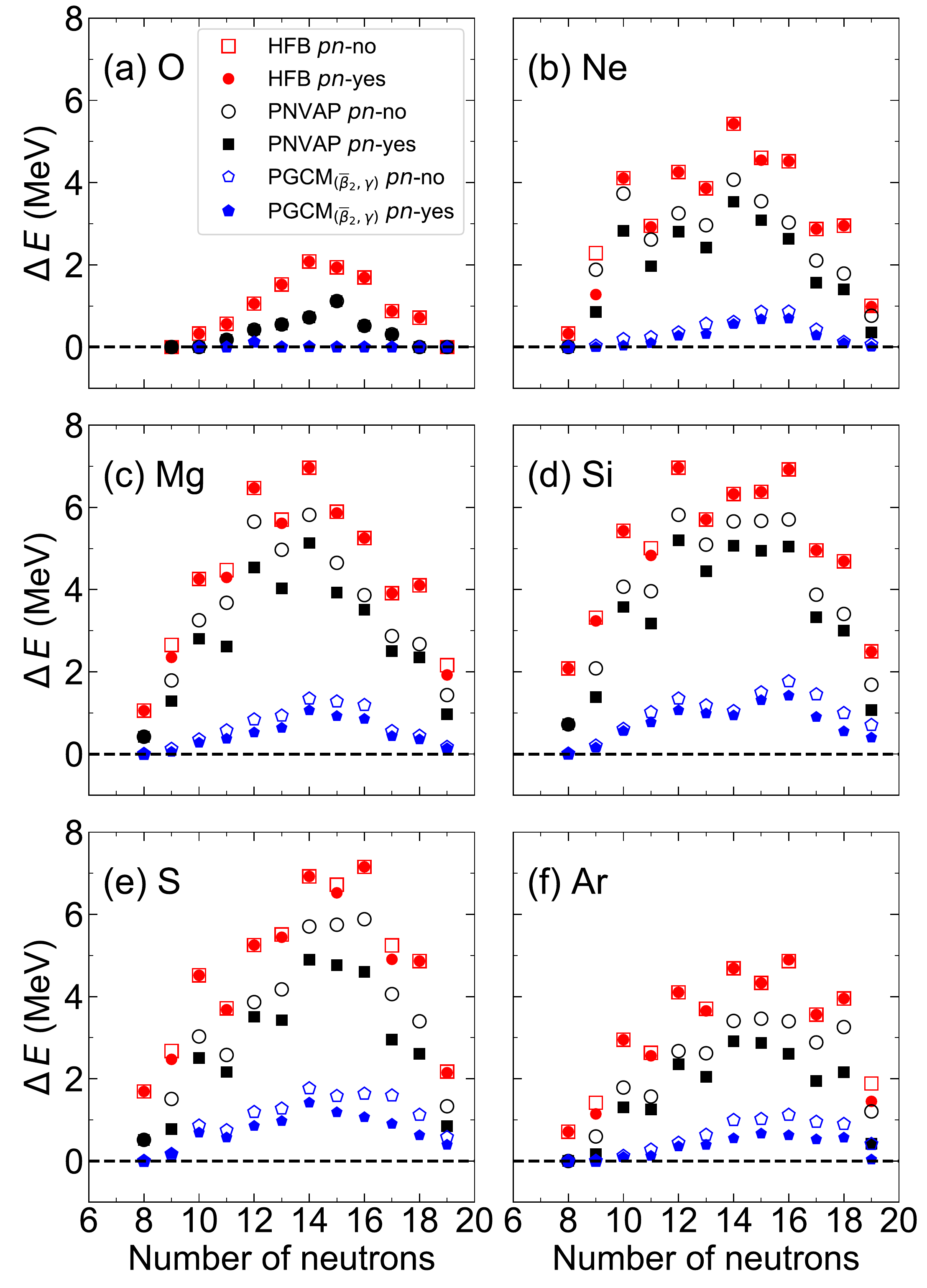}
\end{center}
\caption{(color online) Energy difference with respect to the exact solution for unconstrained HFB (red symbols) and VAPNP (black symbols) calculations and for PGCM calculations using $(\overline{\beta}_{2},\gamma)$ as generating coordinates (blue symbols) for even-even and even-odd nuclei in the $sd$-shell and the USD interaction. Two different choices of Bogoliubov transformations are used: general $pn$-no (boxes and circles) and general $pn$-yes (bullets and filled boxes).}
\label{usd_even_all_gs}
\end{figure}

\begin{figure}[t]
\begin{center}
\includegraphics[width=\columnwidth]{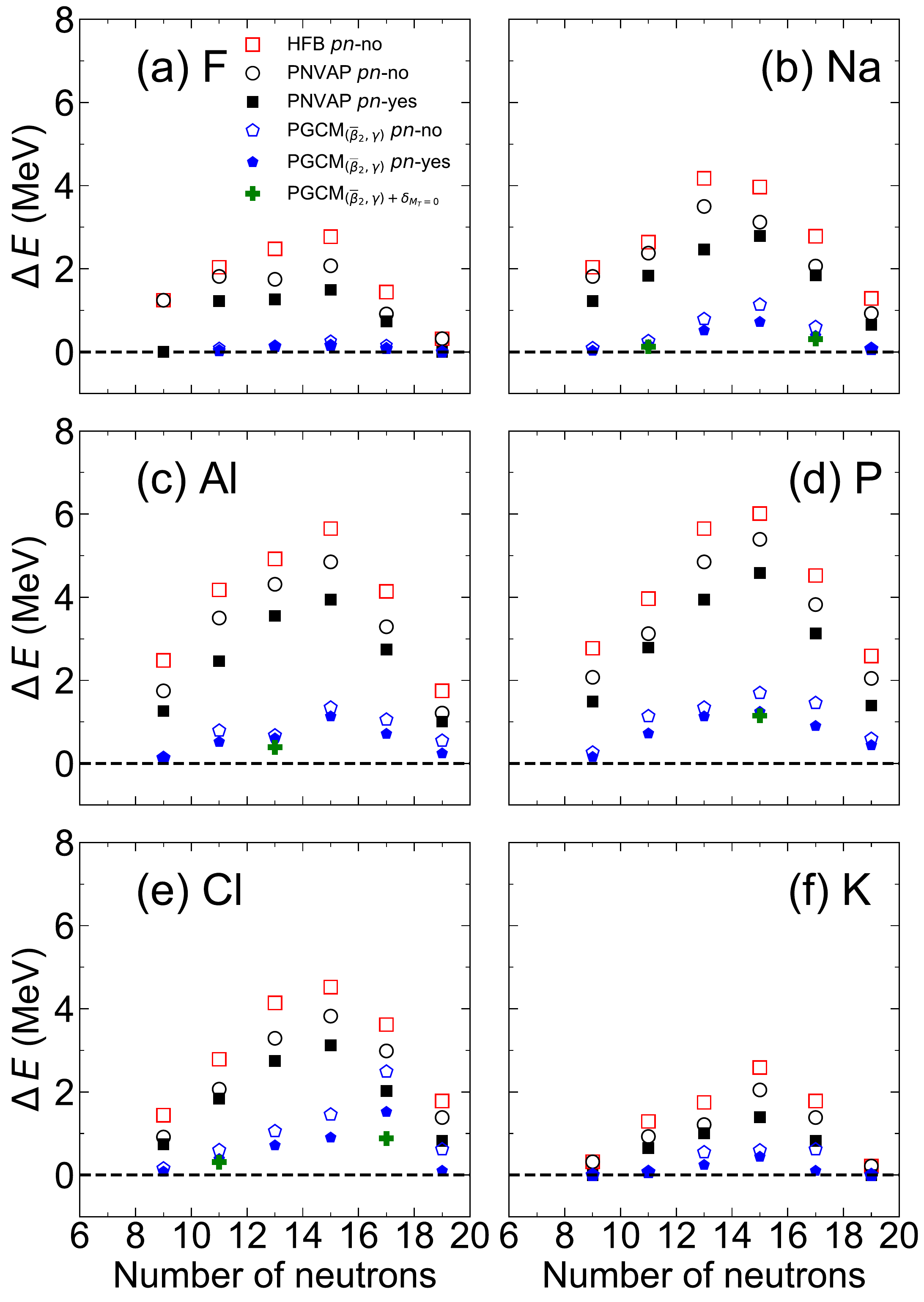}
\end{center}
\caption{(color online) Same as Fig.~\ref{usd_even_all_gs} but for odd-odd isotopes. The HFB results only contains general $pn$-no seeds. The green crosses show the result for PGCM results including $M_{T_p}=0$ pairing degrees of freedom apart from $(\overline{\beta}_{2},\gamma)$ (see text for details).}
\label{usd_odd_odd_gs}
\end{figure}

We extend the unconstrained calculations performed for $^{24}$Ne, $^{25}$Ne and $^{24}$Na to all even-even, even-odd (or odd-even) and odd-odd nuclei in the $sd$-shell. Unless indicated otherwise, the energies will be always substracted by the exact ground state energies obtained from the full diagonalization of the problem, i.e.\ we will discuss the energy differences 
\begin{equation}
\Delta E= E_{\mathrm{approx}}-E_{\mathrm{exact}} .
\end{equation} 
These energy differences are displayed for the various isotopic chains in the $sd$-shell on Fig.~\ref{usd_even_all_gs} (even-even and even-odd nuclei) and Fig.~\ref{usd_odd_odd_gs} (odd-odd nuclei) for the general $pn$-no and $pn$-yes seeds and for both HFB and VAPNP minimizations.
For odd-odd nuclei, we cannot compute the nuclei within the HFB approach with $pn$-mixing because minimizing the energy without forcing the number parity for protons and neutrons to be odd separately produces fake odd-odd quasiparticle vacua that are in reality made of particle-number eigenstates with an even number of particles for both proton and neutron species [see Fig.~\ref{24Na_minimum_nz_dist}(c)]. 

As a general comment on this global calculation, we always obtain positive energy differences. This is not surprising because these approximations miss some correlations contained in the exact eigenstates and because the VAPNP is strictly variational with respect to the space of many-body states with the correct number of particles.
In fact, we observe that the VAPNP method always provides a closer solution to the exact value than the HFB approximation if we use the same kind of trial wave function. The largest differences with the exact solution are found in mid-shell nuclei where the effects of additional symmetry restorations (e.g., angular momentum) and configuration mixing become more important, as it will be illustrated later on. Furthermore, we observe an even-odd staggering superimposed to such a global trend, with the even-odd isotopes being closer to the exact solutions than their even-even neighbors (see Fig.~\ref{usd_even_all_gs}).

Comparing now the solutions with and without including the $pn$-mixing in the variational space, 
we see a different behavior between the HFB and VAPNP approaches. 
Looking first at the HFB case, the differences between those solutions are very small in the even-odd isotopes and even non-existent in the even-even isotopes. If we analyze closely the pairing energies of the even-odd nuclei in the HFB approximation, we observe that in our results all the general $pn$-no solutions have $E^{nn}_{\mathrm{pair}}=0$ (except for $^{19,21}$O) and $E^{pp}_{\mathrm{pair}}=0$ (except for $^{25,27}$Ne, $^{21,31}$Mg, $^{23,25}$Si, $^{25}$S and $^{27,29,37}$Ar). By contrast, all the general $pn$-yes solutions have $E^{pn}_{\mathrm{pair}}\neq0$ (except for the O chain whose solutions are obviously the same as the general $pn$-no ones, and $^{23}$Ne, $^{29}$Mg and $^{27-31}$Si), and $E^{pp/nn}_{\mathrm{pair}}=0$, showing that the $nn/pp$- and $pn$-pairing phases are not mixed in the unconstrained self-consistent mean-field solutions~\cite{PhysRevC.55.1781}. The situation is more extreme in the even-even isotopes where the general $pn$-no and and $pn$-yes solutions are the same because $E^{pn}_{\mathrm{pair}}=0$ is obtained for all nuclei. In particular, in the HFB approximation all $N=Z$ nuclei  collapse to HF solutions with no pairing of any kind.   

The analysis of VAPNP results leads to different conclusions.
Here, a clear difference is observed depending if one allows or not the possibility for protons and neutrons to mix among each others. In particular, the trial wave functions $pn$-yes always give the solution with the lowest energy, except in those cases where there is no proton or neutron in the valence space. It has to be noted that the VAPNP method, which minimizes the particle-number projected energy, always produced paired solutions, i.e.,  with $E_{\mathrm{pair}}\neq0$. More remarkably, the $pn$-yes wave functions always contain at the same time non-vanishing $pp$-, $nn$- and $np-$pairing correlations and that for all nuclei (whether e-e, e-o or o-o systems). This result clearly shows the superiority of the VAPNP approximation that is able to capture more complicated correlations and, therefore, can be used to build better approximations to the eigenstates of the nuclear Hamiltonian.

\subsection{Constrained calculations}
\label{constrained_calc}

\subsubsection{Preliminary study of $^{24}$Ne}

We have analyzed in Sec.~\ref{unconstrained_calc} the ability of the plain HFB and VAPNP methods, which are based on a single intrinsic wave function, 
to approximate the exact solutions. It is clear from the global calculations shown in Figs.~\ref{usd_even_all_gs}-\ref{usd_odd_odd_gs} that non-negligible correlations are still missing. As we will show, most of those correlations can be accounted for by simultaneously restoring the rotational invariance and mixing different quasiparticle states within the PGCM formalism discussed in Sec.~\ref{Theoretical_Framework}. Moreover, this framework permits the calculation of excited states and electromagnetic properties (transition probabilities and moments). 

One of the critical aspects of the method is the selection of the generator coordinates and the natural choices dictated by the nuclear interaction are the quadrupole and pairing degrees of freedom~\cite{Dufour96a}. As an example, we analyze the HFB/VAPNP energy as a function of the quadrupole deformation parameters, $(\overline{\beta}_{2},\gamma)$, and the VAPNP energy as a function of the quadrupole and pairing parameters, $(\overline{\beta}_{2},\delta^{J_p T_p}_{M_{J_p}M_{T_p}})$, for the nucleus $^{24}$Ne. These total energy surfaces (TES) are obtained by solving the HFB/VAPNP equations with constraints (see Sec.~\ref{Theoretical_Framework}).

\begin{figure}[t]
\begin{center}
\includegraphics[width=0.7\columnwidth]{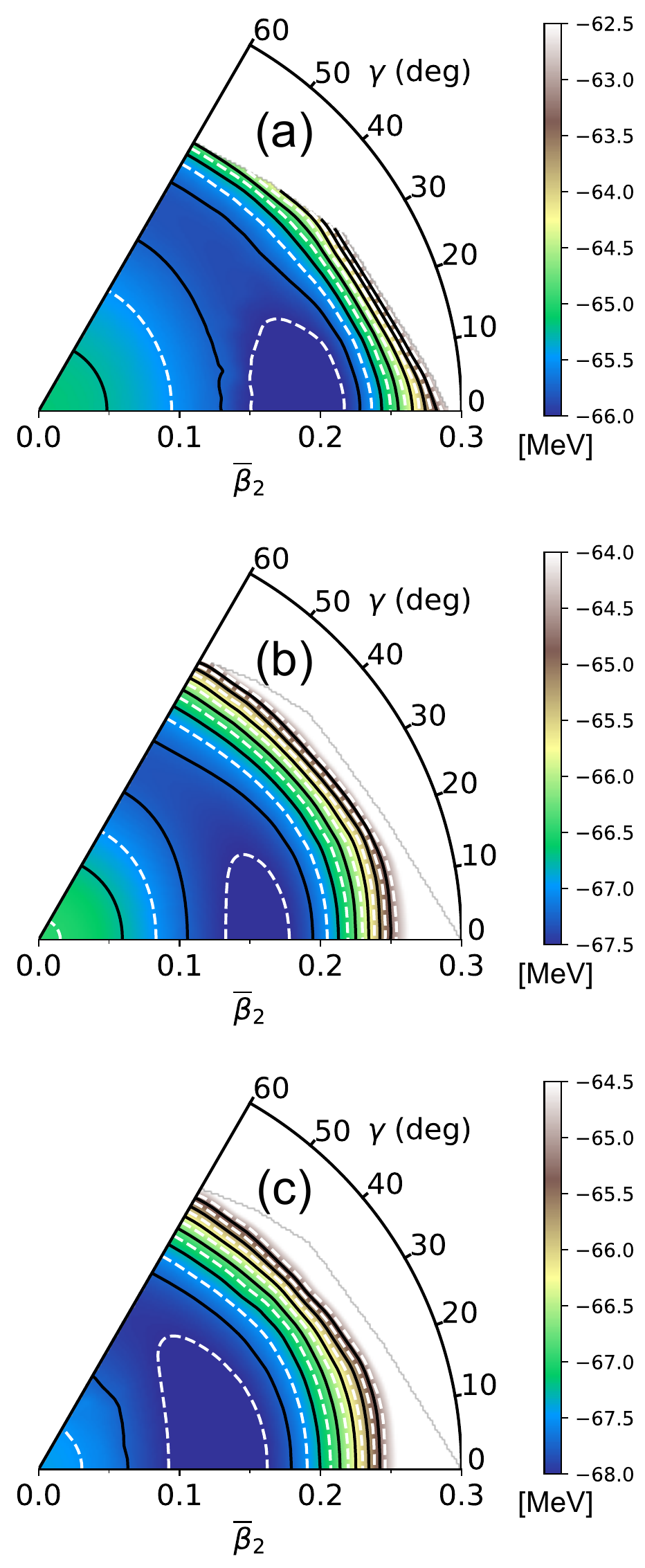}
\end{center}
\caption{(color online) Total energy surfaces as a function of the triaxial deformations $(\overline{\beta}_{2},\gamma)$ calculated for $^{24}$Ne within the following approaches: (a) HFB $pn$-no, (b) VAPNP $pn$-no; and, (c) VAPNP $pn$-yes. Contour lines are separated by intervals of 0.25 MeV and the scale of the color code is different for each plot.}
\label{24Ne_triax_pes}
\end{figure}

We observe in Fig.~\ref{24Ne_triax_pes} that the minima of the TES are found at axial prolate deformations in this isotope both for HFB and VAPNP solutions and the surfaces are softer along the $\gamma$ degree of freedom than in the $\overline{\beta}_{2}$ direction. We also note that the increase of pairing correlations, from HFB to VAPNP $pn$-no and VAPNP $pn$-yes, tends to decrease the deformation. Consistently with the unconstrained results, the VAPNP $pn$-yes method provides the solution with the lowest energy. 

\begin{figure}[htbp]
\begin{center}
\includegraphics[width=\columnwidth]{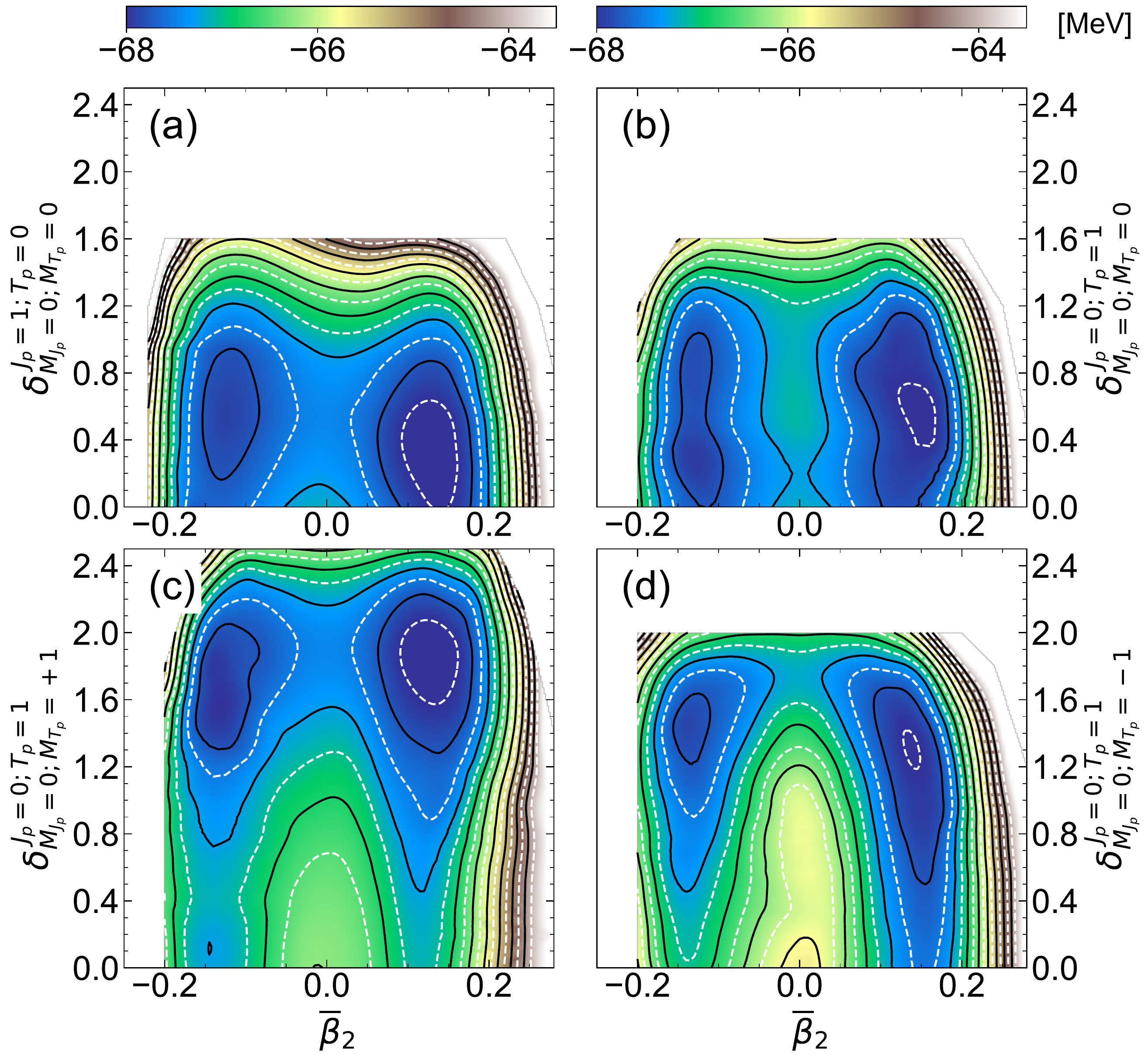}
\end{center}
\caption{(color online) VAPNP total energy surfaces for $^{24}$Ne as a function of the quadrupole deformation, $\overline{\beta}_{2}$, and: (a) isoscalar pairing, $\delta^{J_p=1;T_p=0}_{M_{J_p}=0;M_{T_p}=0}$; (b) isovector $pn$-pairing, $\delta^{J_p=0;T_p=1}_{M_{J_p}=0;M_{T_p}=0}$; (c) $nn$-pairing, $\delta^{J_p=0;T_p=1}_{M_{J_p}=0;M_{T_p}=+1}$; and, (d) $pp$-pairing, $\delta^{J_p=0;T_p=1}_{M_{J_p}=0;M_{T_p}=-1}$. Contour lines are separated by intervals of 0.25 MeV and the scale of the color code is the same in all the plots.}
\label{24Ne_beta_delta_VAPNP}
\end{figure}

Concerning the dependence of the energy on both quadrupole and pairing, we represent in Fig.~\ref{24Ne_beta_delta_VAPNP} the TES calculated with the VAPNP method exploring explicitly the isoscalar $(J_p=1, T_p=0)$ and isovector $(J_p=0, T_p=0)$ pairing degrees of freedom. We point out that increasing the value of $\delta$ in a given channel produces an increase of the corresponding pairing energy, e.g., $\delta^{J_p=0;T_p=1}_{M_{J_p}=0;M_{T_p}=-1}$ measures the amount of $pp$-pairing energy in the intrinsic wave function~\cite{LOPEZVAQUERO2011520,Bally19a}. We obtain in all cases two minima (prolate and oblate) in the TES, the prolate minimum being the absolute one. This is consistent with the triaxial TES discussed above where the oblate minimum is the point with the lowest energy at $\gamma=60^{\circ}$ in Fig.~\ref{24Ne_triax_pes}. The TES along the pairing content, both isoscalar and isovector, are rather soft in the $\delta$ direction around the minima. Additionally, these minima are found at values with $\delta\neq0$, showing the ability of the VAPNP method to include any kind of pairing in the intrinsic wave function. 

\begin{figure}[thbp]
\begin{center}
\includegraphics[width=\columnwidth]{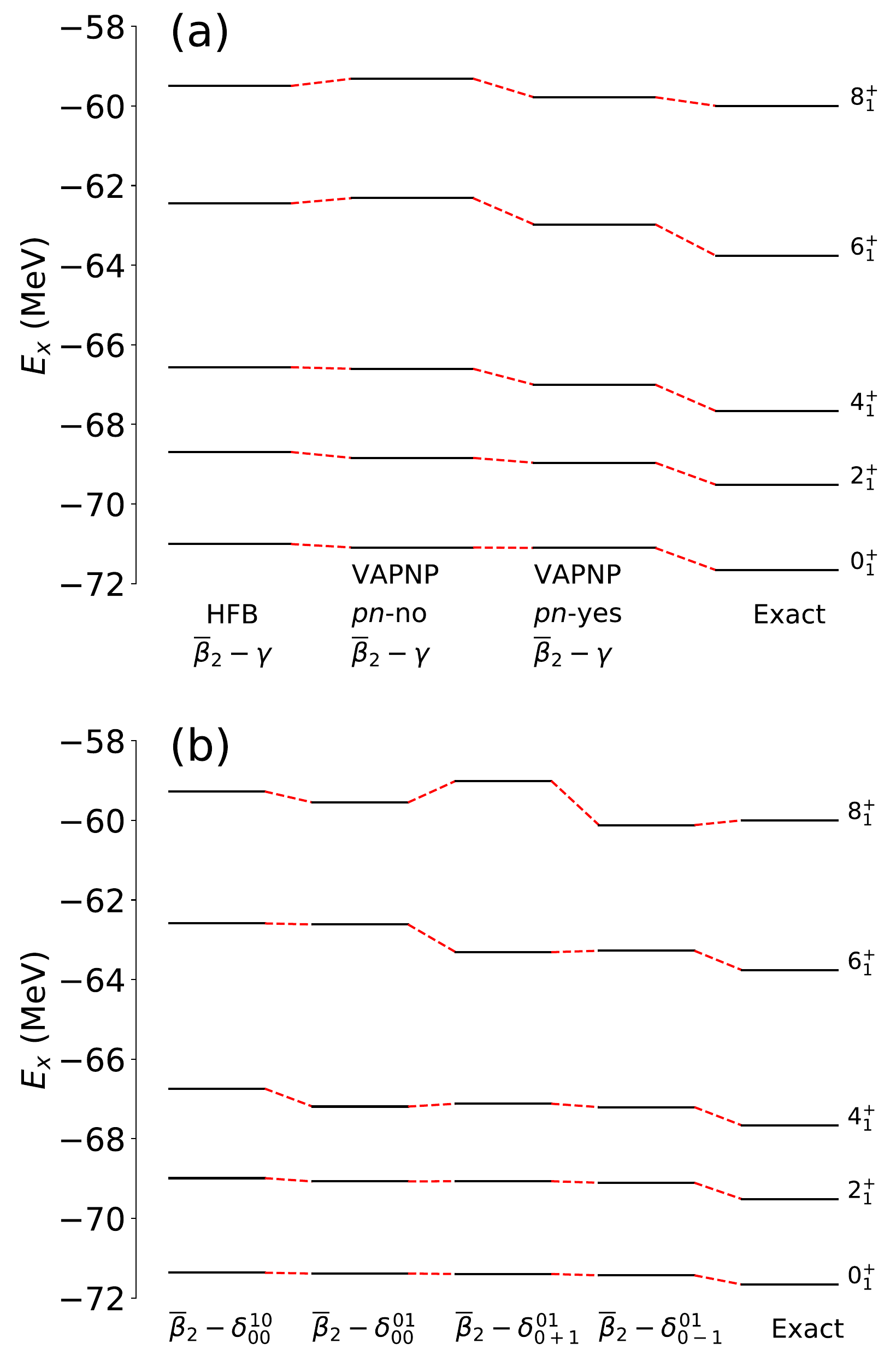}
\end{center}
\caption{(color online) PGCM and exact energies of the yrast states for $^{24}$Ne calculated with the USD interaction. (a) PGCM exploring triaxial $(\overline{\beta}_{2},\gamma)$ deformations; and, (b)  PGCM exploring deformation and pairing $(\overline{\beta}_{2},\delta^{J_p T_p}_{M_{J_p}M_{T_p}})$ degrees of freedom.}
\label{24Ne_spect}
\end{figure}

The softness of the TES shown in Fig.~\ref{24Ne_triax_pes} and Fig.~\ref{24Ne_beta_delta_VAPNP} suggests that the configuration mixing may play an important role in the final results. Therefore, the next step is the calculation of the spectrum of the nucleus $^{24}$Ne within the PGCM framework using the different sets of intrinsic wave functions that define the TES discussed above. The results are represented in Fig.~\ref{24Ne_spect} where we observe a general good agreement between the variational approaches and the exact results. Ground state energies are $1\%$ above the exact energy at most and the best approximations correspond to the PGCM calculations that explore both deformation and pairing explicitly with intrinsic wave functions generated through a VAPNP minimization. We see in Fig.~\ref{24Ne_spect}(a) that the inclusion of intrinsic states with $pn$-mixing compresses the spectrum compared to the two other approaches where the intrinsic states are direct product of separate protons' and neutrons' wave functions (HFB and VAPNP $pn$-no). Nevertheless, we notice that the configuration mixing of HFB states produces good results despite the larger differences found in unconstrained calculations. Concerning the PGCM with deformation and pairing, we observe in Fig.~\ref{24Ne_spect}(b) rather similar results for most of the energies, specially if we compare the results with $pn$-pairing ($\delta_{M_{T_p}=0}$) and the results with $pp/nn$-pairing ($\delta_{M_{T_p}=\pm1}$) among themselves. These similarities are also observed in the evaluation of the electric quadrupole transition probabilities and moments shown in Table~\ref{24Ne_be2}. We have used the standard effective charges, i.e., 1.5 and 0.5 for protons and neutrons, respectively. The PGCM approaches are in a very good agreement with the exact values. For example, all of them are able to reproduce the large [small] $B(E2,2^{+}_{1}\rightarrow0^{+}_{1})$ [$B(E2,2^{+}_{2}\rightarrow0^{+}_{1})$] value, and the sign and magnitudes of the spectroscopic quadrupole moments. Here, the states with a negative spectroscopic moment would be dominated by the prolate configurations shown in the TES above.
\begin{table}[b]
\begin{ruledtabular}
\begin{tabular}{cccccc}
Approx.& $2^{+}_{1}\rightarrow0^{+}_{1}$ & $2^{+}_{2}\rightarrow0^{+}_{1}$ & $2^{+}_{1}$ & $2^{+}_{2}$  &  $4^{+}_{1}$ \\
\hline
$(\overline{\beta}_{2},\gamma)_{\mathrm{HFB}\,pn-\mathrm{no}}$ & 53.3 & 1.0 & -5.4 & +6.5 & -17.2  \\
$(\overline{\beta}_{2},\gamma)_{\mathrm{VAPNP}\,pn-\mathrm{no}}$ &54.1 & 2.1 & -3.8 & +4.8 & -16.8   \\
$(\overline{\beta}_{2},\gamma)_{\mathrm{VAPNP}\,pn-\mathrm{yes}}$ & 54.9 & 2.3 & -4.2 & +5.1 & -16.6  \\\hline
$(\overline{\beta}_{2},\delta^{01}_{00})$ & 53.7 & 2.4 & -3.7 & +2.4 & -15.3 \\
$(\overline{\beta}_{2},\delta^{10}_{00})$ & 53.4 & 1.9  & -4.1 & +3.6 & -14.5  \\
$(\overline{\beta}_{2},\delta^{10}_{-10})$ & 52.9 & 1.4 & -3.8 & +5.3 & -15.4  \\
$(\overline{\beta}_{2},\delta^{10}_{+10})$ & 54.5 & 2.0 & -5.1 & +4.0 & -15.5  \\\hline
Exact & 52.3 & 0.6 & -3.6 & +5.5 & -15.4  \\
\end{tabular}
\end{ruledtabular}
\label{24Ne_be2}
\caption{Electric quadrupole transition probabilities ($B(E2)$ in $e^{2}$ fm$^{4}$) and moments ($Q(J^{\pi})$ in $e$ fm$^{2}$) calculated with different PGCM approximations and with full diagonalization for $^{24}$Ne.}
\end{table}%

\subsubsection{Systematic PGCM calculations}

The results shown above for $^{24}$Ne suggest that any PGCM used to compute the spectrum of this nucleus gives a very good agreement with the full diagonalization of the Hamiltonian. Of course, the degrees of freedom are very limited in this valence space and we expect larger differences between the different PGCM approximations in larger valence spaces or in no-core calculations. Nevertheless, we want to analyze the performance of the method in a more global calculation. Therefore, we compute the energies and electromagnetic properties of the whole $sd$-shell with two PGCM methods whose intrinsic wave functions are obtained with the VAPNP approach using $(\overline{\beta}_{2},\gamma)$ as the generating coordinates (PGCM$_{(\overline{\beta}_{2},\gamma)}$). Since we are particularly interested in the role played by $pn$-mixing, these methods differ in the inclusion or not of the $pn$-mixing in the intrinsic states. 

We analyze first the ground state energies obtained after solving the corresponding HWG equations. The results are shown in Figs.~\ref{usd_even_all_gs}-\ref{usd_odd_odd_gs}. We observe a significant gain in correlation energy with respect to the unconstrained results due to the angular-momentum projection and configuration mixing. Although it is not shown in the plots, the angular momentum of the exact ground state wave functions are reproduced in all even-odd and odd-odd nuclei except for $^{34}$Cl. In this particular case, the exact $0^{+}_{1}$ ground state is almost degenerated with the $3^{+}_{1}$ excited state whereas in the variational approaches, the latter is found to be the ground state and the $0^{+}_{1}$ state is poorly reproduced including only quadrupole collective coordinates. We will come back to that below. 

We see in Figs.~\ref{usd_even_all_gs}-\ref{usd_odd_odd_gs} that the PGCM method with $pn$-mixing is always closer to the exact solution than the PGCM without $pn$-mixing. In the former case, the deviations with respect to the exact g.s. energies are, in general, less than 1 MeV. Similarly to the unconstrained results, these discrepancies are larger for mid-shell nuclei and the degree of agreement with the exact values is similar for even-even, even-odd and odd-odd isotopes. We also see that even-odd staggering is suppressed by performing the configuration mixing. The differences between the two PGCM calculations are smaller than those found in the unconstrained calculations, although, in certain nuclei, such differences could be of the order of 1 MeV (e.g., $^{34}$Cl). But they indicate that including the $pn$-mixing in the intrinsic states can help to better reproduce the ground-state energies. 
Additionally, it is interesting to note that both PGCM approaches perform similarly well regarding the description of the ground state energies of the $N=Z$ nuclei.

A more global and quantitative way of evaluating the ability of the present PGCM methods to reproduce the exact ground state energies is to compute the root-mean-square deviation (RMSD) between the approximate and exact ground-state energies. Such a quantity is displayed in the third column of Table~\ref{rmsd}, separately for 
 e-e, e-o and o-o nuclei and for the $pn$-yes and $pn$-no PGCM calculations. In particular, we see that the former is a better approach than the latter by approximately $250$ keV. Additionally, the deviations in the ground state energies are similar for e-e, e-o and o-o nuclei. The largest RMSD obtained for the PGCM with (without) $pn$-mixing is 628 (871) keV which indicates a very good approximation to the exact results.

\begin{figure*}[htbp]
\begin{center}
\includegraphics[width=\textwidth]{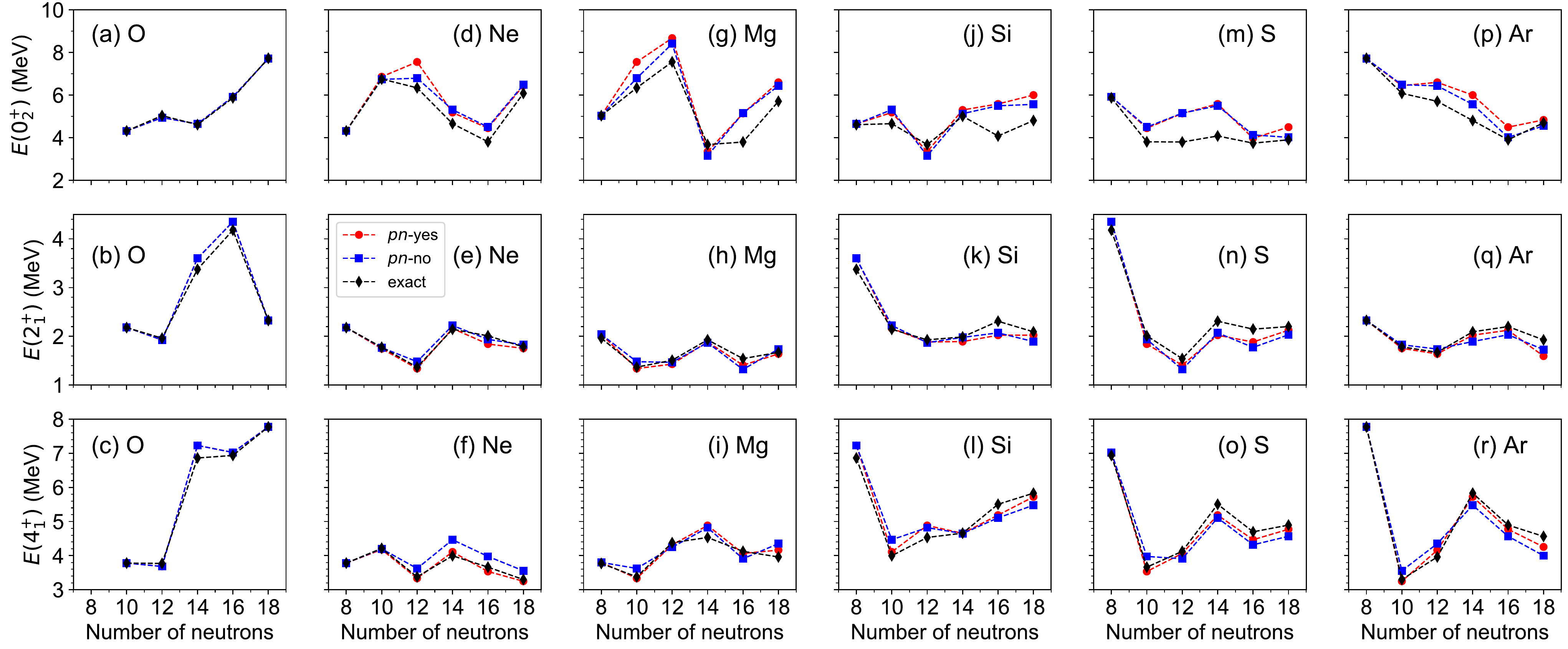}
\end{center}
\caption{(color online) $0^{+}_{2}$ (top panel), $2^{+}_{1}$ (middle panel) and $4^{+}_{1}$ (bottom panel) excitation energies for the even-even isotopes in the $sd$-shell nuclei calculated exactly (black diamonds) and using PGCM with $pn$-mixing (red dots) and PGCM without $pn$-mixing (blue squares) techniques.}
\label{Exc_e_e_all}
\end{figure*}

\begin{figure*}[htbp]
\begin{center}
\includegraphics[width=\textwidth]{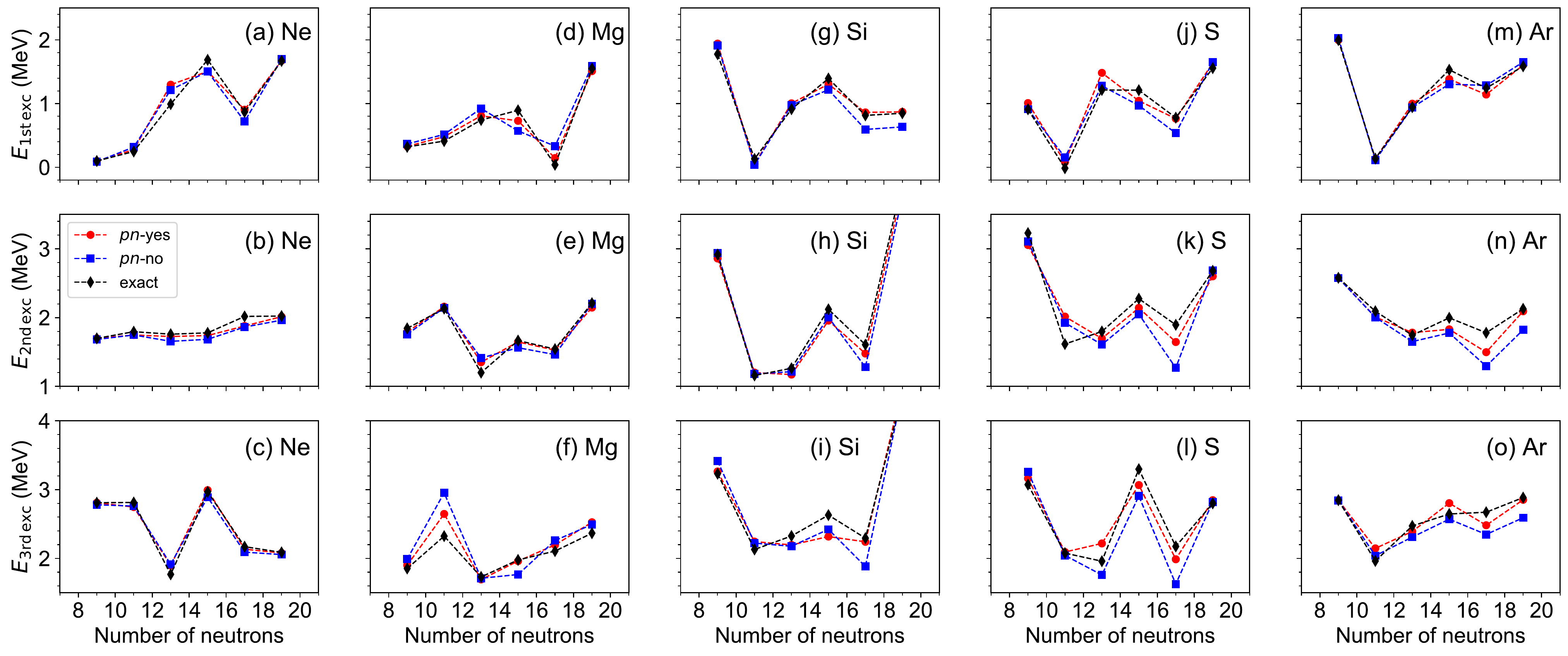}
\end{center}
\caption{(color online) First (top panel), second (middle panel) and third (bottom panel) excitation energies for the even-odd isotopes in the $sd$-shell nuclei calculated exactly (black diamonds) and using PGCM with $pn$-mixing (red dots) and PGCM without $pn$-mixing (blue squares) techniques.}
\label{Exc_e_o_all}
\end{figure*}

\begin{figure}[htbp]
\begin{center}
\includegraphics[width=\columnwidth]{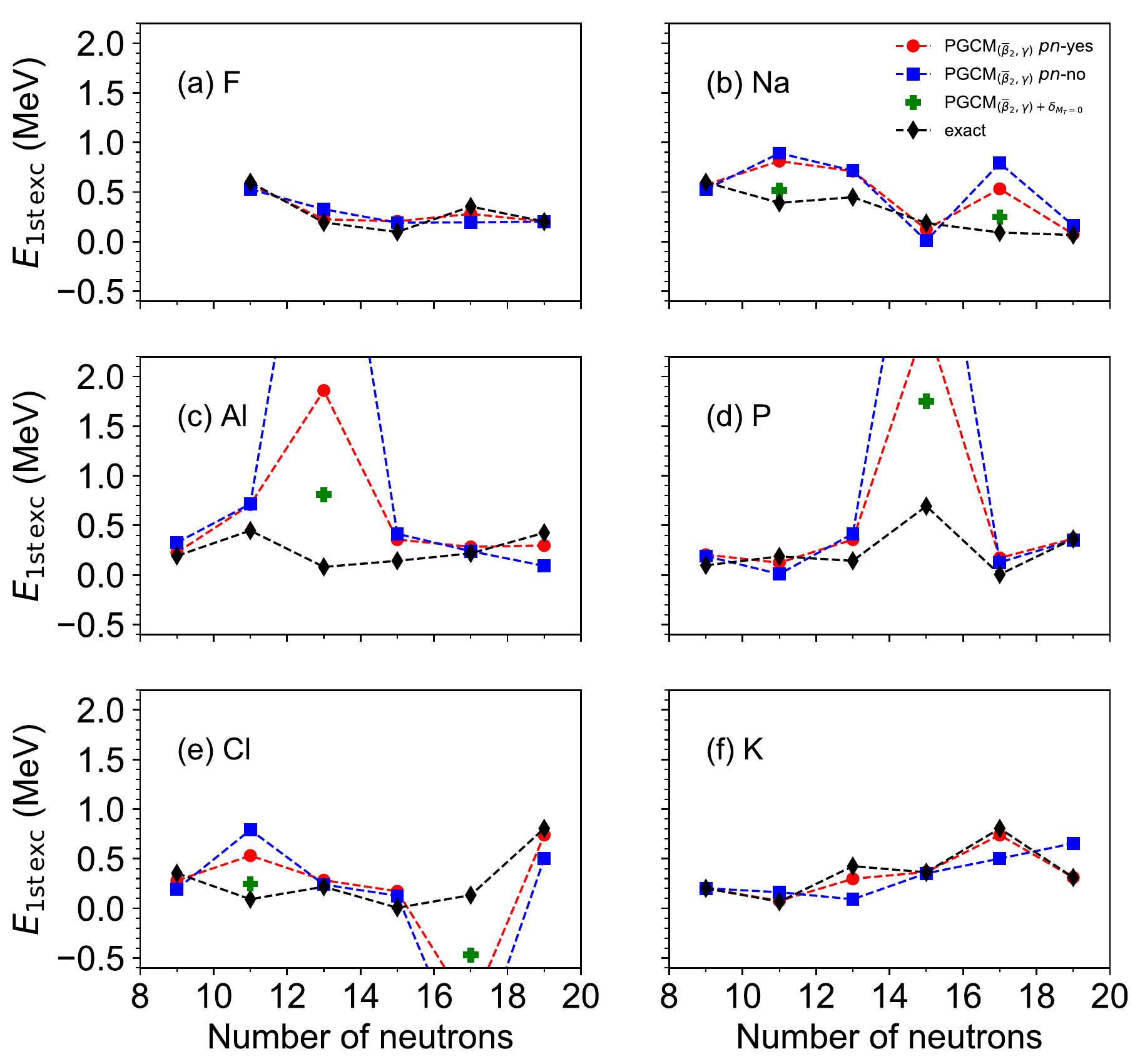}
\end{center}
\caption{(color online) Excitation energies for the first excited states in odd-odd isotopes in the $sd$-shell nuclei calculated exactly (black diamonds) and using PGCM with $pn$-mixing (red dots) and PGCM without $pn$-mixing (blue squares) techniques. The green crosses show the result for PGCM results including $M_{T_p}=0$ pairing degrees of freedom apart from $(\overline{\beta}_{2},\gamma)$ (see text for details)}
\label{Exc_o_o_all}
\end{figure}

We now study the capacity of the PGCM methods to reproduce the excitation energies. We represent the results for the lowest excited states in even-even, even-odd and odd-odd nuclei in Figs.~\ref{Exc_e_e_all},~\ref{Exc_e_o_all} and~\ref{Exc_o_o_all}, respectively. Concerning the even-even isotopes, we observe an excellent reproduction of the exact results for the $2^{+}_{1}$ and $4^{+}_{1}$ excitation energies in all nuclei. The agreement is slightly worse for the $0^{+}_{2}$ excitation energies although the qualitative behavior of this quantity is overall well reproduced, with the exception of the isotopes $^{28}$Mg, $^{28}$S, $^{30}$Si and $^{30}$S where the largest disagreements are observed. In these cases, the explicit exploration of all pairing degrees of freedom and/or the inclusion of quasiparticle excitations in the set of intrinsic states considered would be needed~\cite{LOPEZVAQUERO2011520,Hara_Sun_PSM,PhysRevC.93.064313,PhysRevC.95.024307}.
In the last three columns of Table~\ref{rmsd}, we display the RMSD of the total and excitation energies for the $0^{+}_{2}$, $2^{+}_{1}$, and $4^{+}_{1}$ states. Overall, the $pn$-yes PGCM calculations performs better than $pn$-no PGCM calculations for the total energies as well as for the excitation energies. The only exception are the excitation energies of the $0^{+}_{2}$ states which are slightly worse but this is because the inclusion of $pn$-pairing correlations lowers to a greater extent the ground-state energies than the total energies of the second $0^+$ states. From a more general perspective, we also see that the excitation energies are better reproduced (smaller RMSD values) than the total energies, as expected for these relative quantities.

We also observe a very good agreement with the exact results for even-odd isotopes (Fig.~\ref{Exc_e_o_all}). In this case, we display the excitation energies of the three lowest states obtained from the exact diagonalizations. We have to point out that the values of the angular momentum of the exact excited states are also well reproduced with the variational approaches. As can been seen in  Table~\ref{rmsd}, the RMSD for the total and excitation energies of the first three excited states are slightly smaller when taking into account the $pn$-mixing. This overall better agreement of the  $pn$-yes PGCM calculations is consistent with the results obtained for the e-e nuclei. 
In addition, we point out that the RMSD of the excitation energies of the three lowest excited states are of the order of 100 keV, which is very good.

The main differences are found when looking at the odd-odd isotopes. In Fig.~\ref{Exc_o_o_all}, we show the energies of the first excited state obtained with the exact calculations and their PGCM counterparts. First of all, the angular momentum of both ground and first excited states are the same for the variational approaches and the exact solutions for all odd-odd nuclei, except for the $^{26}$Al, $^{30}$P and $^{34}$Cl nuclei, which are all $N=Z$ isotopes. In the general case, the PGCM excitation energies are in good agreement with the exact ones. But for the aforementioned $N=Z$ isotopes, the main problem comes from the poor description of the $0^{+}_{1}$ state that is the first excited state in $^{26}$Al and $^{30}$P, and the ground state in $^{34}$Cl. For example, the ground state found with the PGCM method exploring the quadrupole degrees of freedom in $^{34}$Cl is the $3^{+}_{1}$ state, which is the first excited state in the exact calculation, while the $0^{+}_{1}$ state has an excitation energy above 1 MeV. Taking this state as the ground state, we then obtain the negative excitation energy represented in Fig.~\ref{Exc_o_o_all}(e) that is only meaningful to compare with exact results. We also observe in Fig.~\ref{Exc_o_o_all}(b)-(e) that the inclusion of $pn$-mixing gives better results in $N=Z$ but this improvement is not sufficient to describe properly these excitation energies. These $0^{+}_{1}$ states are, in fact, the isobaric analogue states with $M_{T}=0$ of the $T=1$ triplet. 

In order to improve the PGCM description of these problematic states, we have performed exploratory calculations including in the set of intrinsic states both states obtained along $(\overline{\beta}_{2},\gamma)$ and along $(\delta^{J_p=0;T_p=1}_{M_{J_p}=0;M_{T_p}=0},\delta^{J_p=1;T_p=0}_{M_{J_p}=0;M_{T_p}=0})$ directions. These calculations have been carried out for the isotopes where the largest deviations with the exact excitation energies are found, i.e., $^{22}$Na, $^{26}$Al, $^{30}$P, $^{34}$Cl, $^{28}$Na and  $^{28}$Cl. The ground state energy gained by adding these pairing degrees of freedom, shown in Fig.~\ref{usd_odd_odd_gs} as green crosses, is almost negligible, except for $^{34}$Cl where around $\sim0.6$ MeV is attained. 
A larger effect is observed in the description of the excitation energies. For $^{22}$Na, $^{28}$Na and $^{28}$Cl the new results are now very close to the exact excitation energies. In these cases, the ground and first excited states are ($3^{+}_{1}$, $1^{+}_{1}$) and ($2^{+}_{1}$, $1^{+}_{1}$) for the former and the latter two, respectively. Hence, the isobaric analogue states are not involved here. For the rest of the $N=Z$ nuclei we also see a significant improvement of the PGCM excitation energies but the results are still far away from the exact data and further correlations should be included. These correlations could be obtained by exploring explicitly quasiparticle excitations. However, since these $0^{+}$ states are $T=1$ states and the USD interaction is isospin conserving, it is very likely that the isospin projection would be the more straightforward way of reproducing these isobaric analogue states. PGCM approximations including such a symmetry restoration will be explored in the future but is beyond the scope of the present work. Finally, looking at the RMSD of the total and excitation energies for the three lowest excited states in the o-o nuclei, shown in Table~\ref{rmsd}, we see that the performance of the PGCM method without $pn$-mixing is significantly worse than the PGCM with $pn$-mixing.

\begin{figure}[htbp]
\begin{center}
\includegraphics[width=\columnwidth]{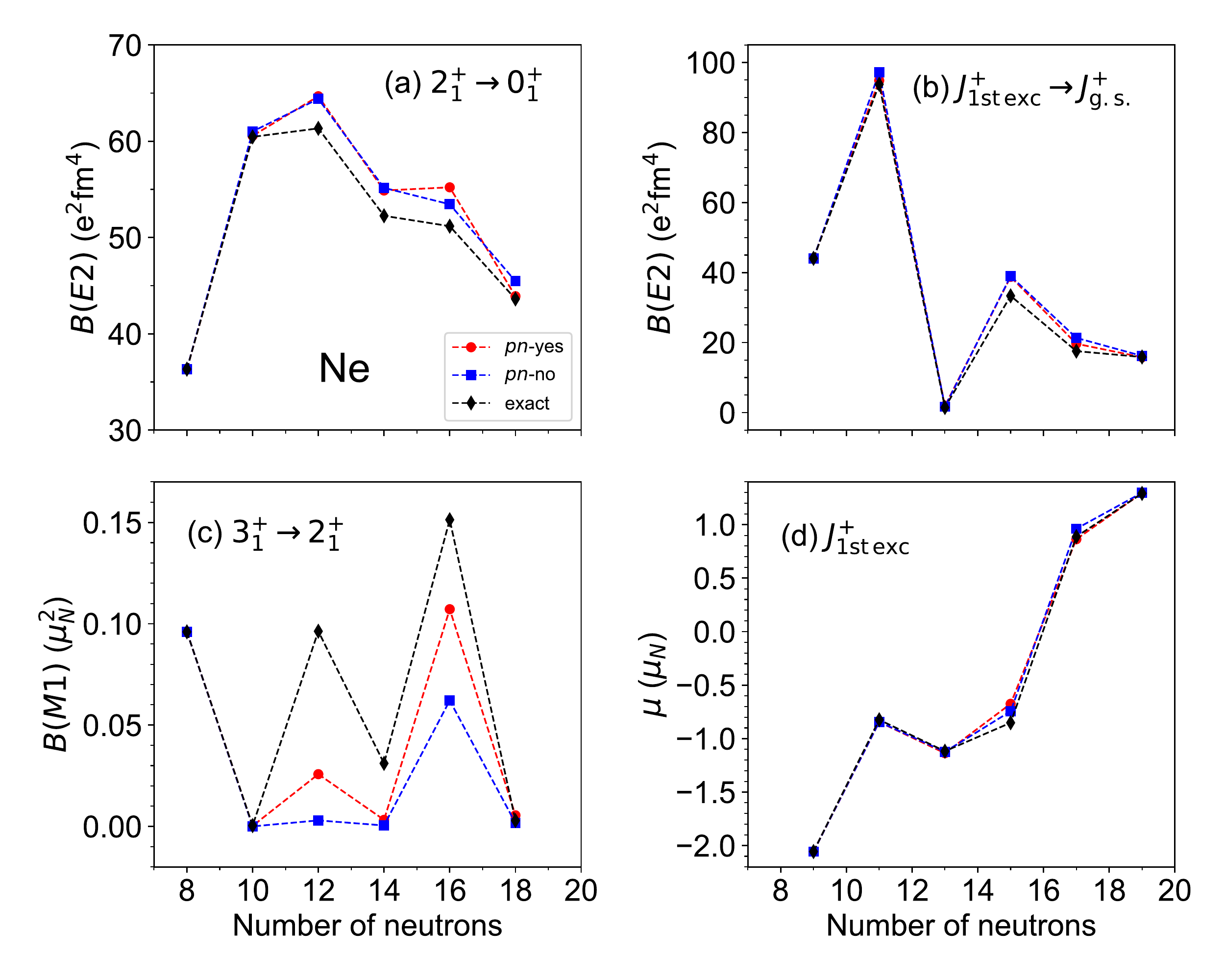}
\end{center}
\caption{(color online) Electromagnetic properties calculated with PGCM with (red bullets) and without (blue squares) $pn$-mixing, and the exact values (black diamonds) for selected states in Ne isotopes: (a) $B(E2,2^{+}_{1}\rightarrow0^{+}_{1})$) in even-even isotopes; (b) $B(E2,J^{+}_{\mathrm{1st\,exc.}}\rightarrow J^{+}_{\mathrm{g.s.}})$) in even-odd isotopes; (c) $B(M1,3^{+}_{1}\rightarrow2^{+}_{1})$) in even-even isotopes; (d) magnetic dipole moment for the first excited state in even-odd isotopes.}
\label{Ne_em}
\end{figure}

\begin{table}[hbt]
\centering
\begin{tabular}{c|c|c|c|c|c}
\hline
\hline
Nuclei & mixing & g.s. & $0_2^+$ &  $2_1^+$ &  $4_1^+$ \\
\hline
$e-e$ & $pn-$yes & 0.628 & 1.277 (0.746) &  0.646 (0.137) & 0.610 (0.178)\\
           & $pn-$no & 0.853 &1.366 (0.631)& 0.759 (0.148) & 0.831 (0.294)\\
\hline
\hline
Nuclei & mixing & g.s. & $1^{st}$ exc. &  $2^{nd}$ exc. &  $3^{rd}$ exc. \\
\hline
$e-o$ & $pn-$yes & 0.564 & 0.571 (0.106) & 0.604 (0.123) & 0.565 (0.128)\\
           & $pn-$no & 0.793 &0.751 (0.144) & 0.690 (0.191) & 0.716 (0.211)\\
\hline
\hline
$o-o$ & $pn-$yes & 0.585 & 0.878 (0.504) & 0.721 (0.314) & 0.730 (0.419)\\
           & $pn-$no & 0.871 & 1.580 (1.153) & 1.190 (0.865) & 0.946 (0.536)\\
\hline
\hline
\end{tabular}
\caption{Root-mean-square energy deviation (in MeV) of the PGCM variational approaches (with and without $pn$-mixing) with respect to the exact results for even-even, even-odd and odd-odd nuclei. For the excited states, the left and right numbers indicate the deviation with respect to the values of the absolute energy and the excitation energy, respectively. }
\label{rmsd}
\end{table}
Lastly, we look at the description of the electromagnetic properties and take the Ne isotopic chain as an illustrative example. In Fig.~\ref{Ne_em},
we show the $B(E2,2^{+}_{1}\rightarrow0^{+}_{1})$ and $B(M1,3^{+}_{1}\rightarrow2^{+}_{1})$ for the even isotopes and the $B(E2,J^{+}_{\mathrm{1st\,exc}}\rightarrow J^{+}_{\mathrm{g.s.}})$ and $\mu(J^{+}_{\mathrm{g.s.}})$ for the odd ones. These quantities are calculated with the usual effective charges (1.5 and 0.5 for protons and neutrons, respectively) and bare nucleon $g$-factors. Similarly to the excitation energies discussed above, the agreement of the PGCM approaches with the exact values is very good, especially for the odd isotopes. We do not find large differences between the results obtained including or not the $pn$-mixing, except for the $B(M1)$ where the PGCM with $pn$-mixing is better. The trends of the exact results are well reproduced although the PGCM values are systematically larger [smaller] for the $B(E2)$ [$B(M1)$]. These discrepancies could be reduced by describing better the excited state through the inclusion of additional degrees of freedom on top of the triaxial deformations $(\overline{\beta}_{2},\gamma)$ (e.g., cranking~\cite{BORRAJO2015341,PhysRevC.103.024315}). 

\section{Summary}
\label{Summary}

In this article, we evaluated the merits of several variational methods by systematically
comparing their results to the exact ones across the $sd$-shell valence space. In particular, we studied not only even-even isotopes but also even-odd and odd-odd nuclei, the latter two being often neglected in variational approximations based on the use of Bogoliubov vacua. For this study, we considered the well-known USD nuclear interaction \cite{USD_BROWN} that has the double advantage of being highly non-trivial and reproducing well experimental data. The advanced variational calculations were performed with the solver TAURUS \cite{TAURUS1} whereas the exact results were obtained by using the shell model code ANTOINE \cite{ant3}.

To better understand the role of the various correlations at play in nuclear systems, we considered a diverse set of variational methods ranging from the plain HFB approach to the better VAPNP scheme to several variants of the more elaborate PGCM that includes beyond-mean-fields correlations through the restoration of the broken symmetries and the mixing of configurations. Also, in each case we considered several types of trial Bogoliubov vacua that differ in their conserved intrinsic symmetries. In particular, we compared the differences between using vacua that include or not a mixing of the proton and neutron single-particle states. To determine the importance of the $pn$-mixing could be of great interest to improve the variational approaches based on an energy density functional as this degree of freedom is almost always neglected in this context. Finally, we want to stress that the even and odd number parity Bogoliubov vacua were treated on the same footing, i.e.\ with a self-consistent solving of the appropriate HFB and VAPNP equations.

The main conclusions that we can draw from this study are the following:
\begin{itemize}
    \item In addition to being better from a variational point of view, general Bogoliubov vacua that break all possible symmetries have the immense practical advantage that they converge much more frequently towards the absolute minimum. This greatly simplifies the work of the practitioner as the calculations are much easier to carry out. 
    
    \item The Bogoliubov vacua that include $pn$-mixing can be used to describe odd-odd systems at the VAPNP level without the need to perform an explicit blocking of two quasiparticles. By contrast, HFB calculations of odd-odd nuclei with $pn$-mixing produced fake odd-odd quasiparticle vacua, i.e.\ vacua that contain only even-even particle-number eigenstates when projected onto good $N$ and $Z$, and thus were not considered. 
    
    \item One of the advantage of the VAPNP scheme is its much better treatment of the pairing correlations without the need to perform additional constraint during the minimization procedure. Importantly, the VAPNP prevents the collapse of the pairing correlations often observed in HFB calculations in situations where the effective pairing interaction is small (e.g., shell closures or in odd systems). Also, when allowing the mixing of protons and neutrons, the VAPNP scheme is able to include simultaneously isovector ($pp$, $nn$ and $pn$) and isoscalar ($pn$) pairing correlations in the intrinsic states. By contrast, in (unconstrained) HFB calculations only one type of pairing correlations survives in the solution.
    
    \item The variational methods relying on the VAPNP minimization and exploring the space of Bogoliubov vacua with $pn$-mixing give the best results for the ground-state and excited-state energies as well as for the electromagnetic properties. In particular, the best variational method in this study was the PGCM using the triaxial deformations as collective coordinates and intrinsic states with $pn$-mixing. We note, however, that the effect of including $pn$-mixing in the description is less important in the PGCM calculations than in the pure VAPNP calculations based on a single intrinsic state.
    
    \item The PGCM is able to reproduce very well the exact results even in the cases of mid-shell nuclei where the Hilbert space of the interacting particles has a larger dimension. This good agreement is observed for the ground-state and excited-state energies as well as for the electromagnetic properties. Nevertheless some discrepancies appear for some energies and transitions and for certain $N=Z$ odd-odd nuclei. Concerning the latter, one major problem is the poor description of the $0^{+}_{1}$ isobaric analogue states. Exploratory calculations using the pairing channels as a generator coordinates show an improvement in the description of those nuclei. But since these are pure $T=1$ states, we expect that including a full isospin projection 
    in the construction of the PGCM states will provide us with a better approximation to those states and some work along this line is in progress.
\end{itemize}

Overall, this study shows the ability of sophisticated variational methods to reproduce exact results of a realistic Hamiltonian in a small valence space, namely, the USD interaction in the $sd$-shell. Nevertheless, some discrepancies exist for specific states or nuclei. This motivates us to considering even more advanced variational calculations that include, e.g., quasiparticle excitations, cranking, or isospin projection. 

\section*{Acknowledgements}

We would like to thank A. Poves, F. Nowacki and L. M. Robledo for useful discussions. This project has received funding from the European Union’s Horizon 2020 research and innovation programme under the Marie Skłodowska-Curie grant agreement No. 839847. The work of TRR was supported by the Spanish MICINN under PGC2018-094583-B-I00, and, in part, by the ExtreMe Matter Institute EMMI
at the GSI-Darmstadt, Germany.  We acknowledge the computer resources and assistance provided by GSI-Darmstadt, TU-Darmstadt  and Centro de Computación Científica-Universidad Autónoma de Madrid (CCC-UAM) computing facilities.



%

\end{document}